\documentclass[a4paper,fleqn]{cas-sc}

\usepackage[numbers]{natbib}
\usepackage{amsmath,amsfonts}
\usepackage{array}
\usepackage[caption=false,font=normalsize,labelfont=sf,textfont=sf]{subfig}
\usepackage{textcomp}
\usepackage{stfloats}
\usepackage{url}
\usepackage{verbatim}
\usepackage{graphicx}
\usepackage{enumitem}
\usepackage{multirow}
\usepackage{booktabs}
\usepackage{makecell}
\usepackage{tablefootnote}
\usepackage{diagbox}
\usepackage{hhline}
\usepackage{pifont}
\usepackage{threeparttable}
\usepackage{xstring}
\usepackage{tabularray}
\usepackage[ruled,linesnumbered]{algorithm2e}
\usepackage[title]{appendix}

\def\tsc#1{\csdef{#1}{\textsc{\lowercase{#1}}\xspace}}
\tsc{WGM}
\tsc{QE}

\newcommand{\plevel}[1]{%
\IfEqCase{#1}{%
{1}{$\triangle$}%
{2}{$\triangle\triangle$}%
{3}{$\triangle\triangle\triangle$}%
}[]%
}%
\newcommand{\plevelstar}[1]{%
\IfEqCase{#1}{%
{1}{$\star$}%
{2}{$\star\star$}%
{3}{$\star\star\star$}%
}[]%
}%

\makeatletter
\newcommand{\ssymbol}[1]{^{\@fnsymbol{#1}}}

\begin{document}
\let\WriteBookmarks\relax
\def\floatpagepagefraction{1}
\def\textpagefraction{.001}

\shorttitle{Affine Modulation-based Audiogram Fusion Network for Joint Noise Reduction and Hearing Loss Compensation}    

\shortauthors{Ni et~al.}  

\title [mode = title]{Affine Modulation-based Audiogram Fusion Network for Joint Noise Reduction and Hearing Loss Compensation} 



%

\author[1]{Ye Ni} [orcid=0009-0002-4411-3325]
\ead{niye@seu.edu.cn}

\author[1,2]{Ruiyu Liang} [orcid=0000-0002-6813-4203]
\cormark[1]
\ead{101101220@seu.edu.cn}

\author[3]{Xiaoshuai Hao}[orcid=0009-0007-4209-6695]
\cormark[1]
\ead{haoxiaoshuai@xiaomi.com}

\author[1]{Jiaming Cheng} [orcid=0000-0002-7136-7876]
\ead{230198469@seu.edu.cn}

\author[2]{Qingyun Wang} [orcid=0000-0002-5000-2966]
\ead{wangqingyun@njit.edu.cn}

\author[1]{Chengwei Huang} [orcid=0000-0001-9060-6361]
\ead{huangcwx@126.com}

\author[1]{Cairong Zou} [orcid=0009-0005-0008-5272]
\ead{cairong@seu.edu.cn}

\author[4]{Wei Zhou} 
\ead{zhouw26@cardiff.ac.uk}

\author[5]{Weiping Ding} 
\ead{dwp9988@163.com}

\author[6,7]{Bj\"orn W.\ Schuller} [orcid=0000-0002-6478-8699]
\ead{schuller@tum.de}

\affiliation[1]{organization={School of Information
Science and Engineering, Southeast University},
city={Nanjing},
postcode={210096}, 
country={China}}

\affiliation[2]{organization={School of Communication Engineering, Nanjing Institute of Technology},
city={Nanjing},
postcode={211167}, 
country={China}}

\affiliation[3]{organization={Xiaomi EV},
	addressline={Xiaomi Campus, Anningzhuang Road, Haidian District}, 
	postcode={100085}, 
	state={Beijing},
	country={China}}
            
\affiliation[4]{organization={Cardiff University},
	city={Cardiff},
	country={United Kingdom}}
	
\affiliation[5]{organization={School of Artificial Intelligence and Computer Science, Nantong University},
city={Nantong},
postcode={226019}, 
country={China}}


\affiliation[6]{organization={CHI — Chair of Health Informatics, Technical University of Munich University Hospital},
city={Munich},
postcode={81675}, 
country={Germany}}

\affiliation[7]{organization={GLAM — Group on Language, Audio, \& Music, Imperial College London},
city={London},
postcode={SW7 2AZ}, 
country={United Kingdom}}

\cortext[1]{Corresponding authors.}



\begin{abstract}
Hearing aids (HAs) are widely used to provide personalized speech enhancement (PSE) services,  improving the quality of life for individuals with hearing loss. 
However, HA performance significantly declines in noisy environments as it treats noise reduction (NR) and hearing loss compensation (HLC) as separate tasks.
This separation leads to a lack of systematic optimization, overlooking the interactions between these two critical tasks, and increases the system complexity.
To address these challenges, we propose a novel audiogram fusion network, named AFN-HearNet, which simultaneously tackles the NR and HLC tasks by fusing cross-domain audiogram and spectrum features.
We propose an audiogram-specific encoder that transforms the sparse audiogram profile into a deep representation, addressing the alignment problem of cross-domain features prior to fusion.
To incorporate the interactions between NR and HLC tasks, we propose the affine modulation-based audiogram fusion frequency-temporal Conformer that adaptively fuses these two features into a unified deep representation for speech reconstruction.
Furthermore, we introduce a voice activity detection auxiliary training task to embed speech and non-speech patterns into the unified deep representation implicitly.
We conduct comprehensive experiments across multiple datasets to validate the effectiveness of each proposed module.
The results indicate that the AFN-HearNet significantly outperforms state-of-the-art in-context fusion joint models regarding key metrics such as HASQI and PESQ, achieving a considerable trade-off between performance and efficiency.
The source code and data will be released at https://github.com/deepnetni/AFN-HearNet.
\end{abstract}


\begin{highlights}
\item Linear interpolation provides a more suitable approach for extending the resolution of audiograms.
\item Affine modulation fusion outperforms the in-context fusion strategy for audiogram and spectrum integration.
\item Voice activation probability information contributes to personalized speech enhancement in hearing aids.
\end{highlights}


\maketitle

\section{Introduction}
\label{intro}

According to the World Health Organization, by 2050, over 700 million people—around one in ten—are projected to experience disabling hearing loss.
If left untreated, hearing loss can lead to various complications, including social isolation and even dementia.
In this context, hearing aids (HAs) play a crucial role in improving the hearing-specific health-related quality of life for individuals with hearing loss~\cite{ferguson2017hearing}.
By enhancing understanding during conversations, HAs enable users to connect more effectively with family, friends, and colleagues. 
This personalized speech enhancement (PSE) service not only improves communication but also helps reduce the risk of associated issues, fostering better overall well-being.

\begin{figure*}[t]
	\setlength{\abovecaptionskip}{-0.2cm}
	\centering
	\includegraphics[width=\linewidth]{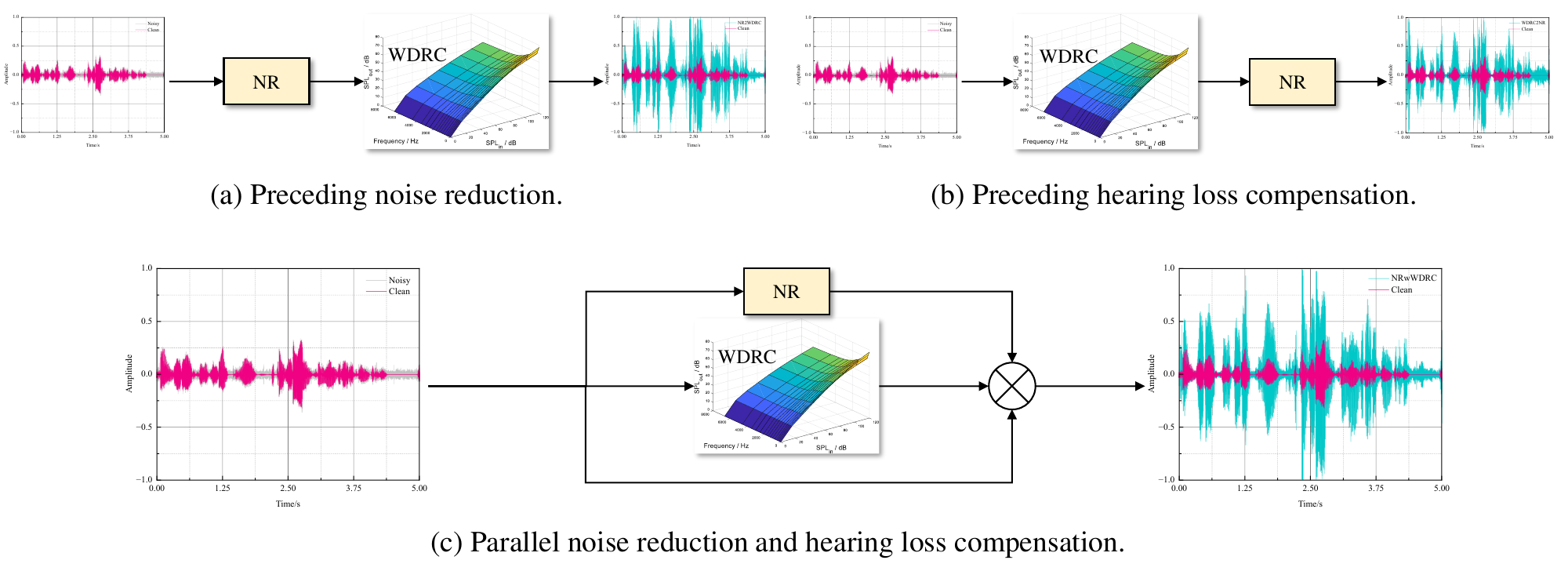}
	\caption{Examples of NR and HLC processes in hearing aids include: serial concatenation, such as (a) preceding NR and (b) preceding HLC, each followed by another component; and (c) parallel NR and HLC, where both NR and the WDRC algorithm are applied independently to the noisy speech.} 
	\label{fig_example}
	\vspace{-0.2cm}
\end{figure*}
\begin{figure}[t]
	\setlength{\abovecaptionskip}{-0.2cm}
	\centering
	\includegraphics[width=0.5\linewidth]{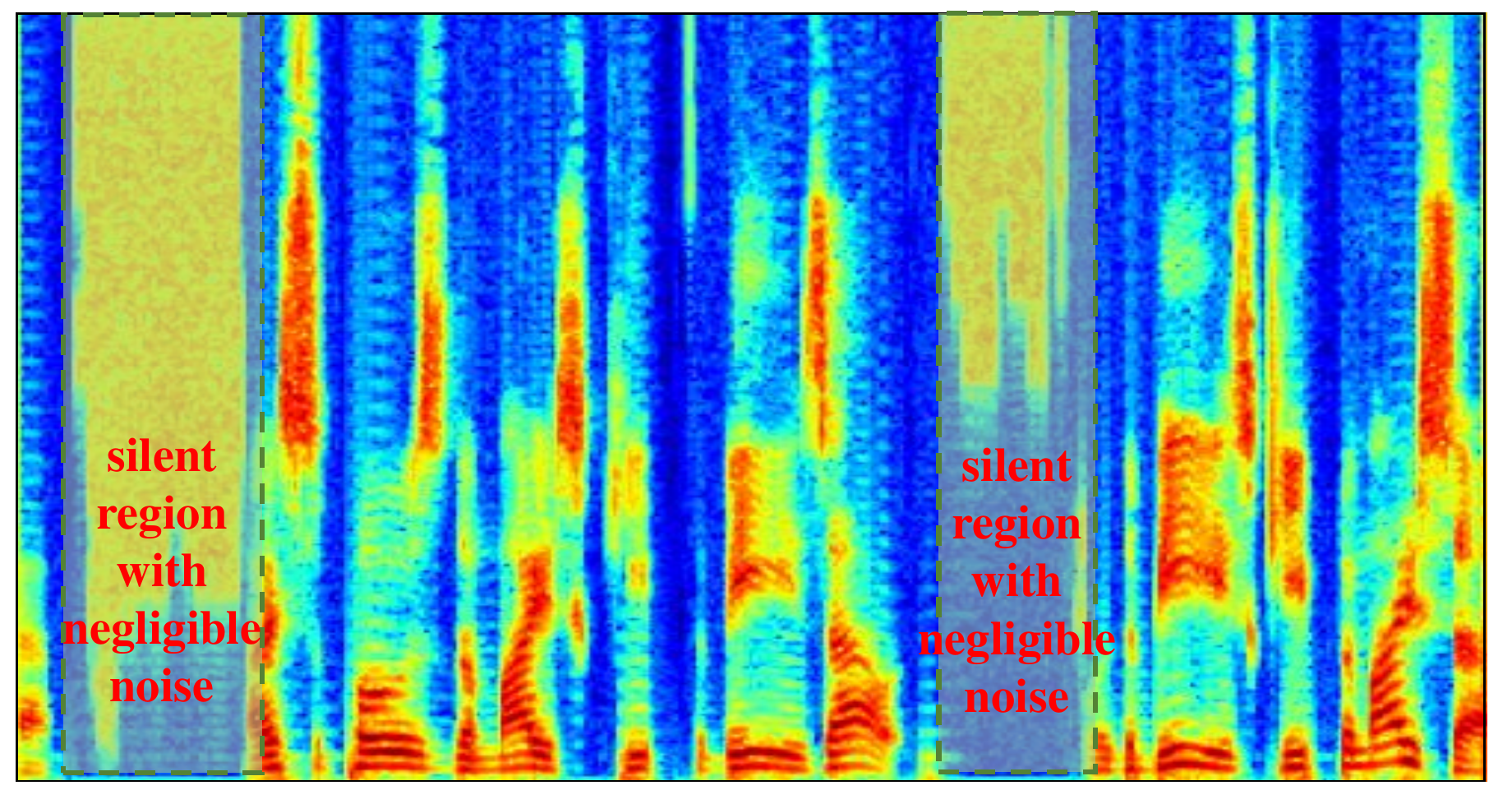}
	\caption{Example of a compensated speech spectrogram, where dashed boxes indicate silent regions with negligible noise.}
	\label{fig_example_2}
	\vspace{-0.2cm}
\end{figure}

To restore audibility, HAs commonly employ hearing loss compensation (HLC), a nonlinear amplification strategy that reshapes the acoustic input to accommodate the limited dynamic range of listeners with hearing impairment.
The most widely used approach is the wide dynamic range compression (WDRC) algorithm, which mimics the function of outer hair cells in the cochlea by providing level-dependent amplification, thereby improving audibility while maintaining loudness comfort.
Despite being the most effective means of hearing rehabilitation, HLC inevitably amplifies not only speech but also background noise, which substantially degrades listening comfort and may impose additional perceptual burdens on hearing-impaired listeners under noisy conditions.
To alleviate this issue, HAs typically utilize a noise reduction (NR) component to suppress background noise prior to amplification. Fig.\,\ref{fig_example} illustrates the existing topologies of NR and HLC modules in HAs. 
Although traditional NR algorithms~\cite{abd2008speech,madhu2012potential,abd2014speech} are very robust when dealing with stationary noise, their performance is degraded in a nonstationary noisy environment~\cite{jannu2023weibull}.
Whether configured in a series or parallel manner, the enhanced speech often retains considerable residual interference, which reveals the inherent limitations of traditional HA frameworks in adapting to complex acoustic conditions.
As illustrated in Fig.\,\ref{fig_example_2}, even when residual noise seems negligible, the WDRC algorithm can still induce pronounced spectral distortions, particularly in non-speech segments and within the high-frequency range.
Consequently, HAs often perform poorly in noisy environments~\cite{souza2006measuring, corey2021modeling}, where individuals with hearing loss require the most assistance.

Recently, deep neural networks (DNN) models utilizing the frequency-temporal sequence modules, e.g., FT-LSTM~\cite{luo2020dual, le2021dpcrn} and FT-Conformer~\cite{cao2022cmgan}, have emerged as powerful backbones for speech enhancement, providing a promising option to address NS and HLC tasks.
Building on these advances, considerable efforts have been dedicated to developing DNN-based systems for HAs~\cite{tu2021two,graetzer2021clarity,akeroyd20232nd,cornell2023multi}, focusing on optimizing NR, HLC, or both components. 
In this line, DNN-based NR models~\cite{schroter2022low, tammen2022deep, westhausen2024real, lei2023low} have demonstrated strong performance in enhancing speech clarity under noisy conditions, while DNN-based HLC models~\cite{drgas2023dynamic, leer2025hearing} have achieved notable success in improving the auditory experience of individuals with hearing loss.
However, these approaches typically focus on independently optimizing a single module, thereby overlooking the crucial interplay between NR and HLC tasks, while the absence of systematic joint optimization further constrains their overall effectiveness.
Various studies~\cite{ngo2008integrated,ngo2012combined,morvan2024factors} have confirmed that simply concatenating NR and HLC components in series remains inadequate.
This limitation highlights the need for a unified framework that can jointly model both processes.

In this work, we propose the affine modulation-based audiogram fusion network (AFN-HearNet), which simultaneously performs NR and HLC by fusing audiogram and spectral features within a unified architecture.
One of the major challenges lies in the effective integration of audiogram information into the speech spectrum representation, which is essential for jointly addressing enhancement and compensation tasks.
While spectral features are dynamic, high-dimensional, and time-dependent, the audiogram is a static, low-dimensional profile that reflects hearing loss levels at discrete frequency bands but lacks temporal structure.
Current studies~\cite{cheng2023speech,yang2025non} address the resolution gap by extending audiograms through repeat-embedding methods.
Fletcher–Munson equal-loudness contours~\cite{fletcher1933loudness}, along with subsequent psychoacoustic studies, indicate that human hearing sensitivity varies smoothly between standard audiometric test frequencies.
This smoothness assumption provides a principled basis for reconstructing intermediate frequency thresholds through linear interpolation.
Building on this insight, we propose an audiogram-specific encoder that transforms the sparse hearing loss profile into a deep representation, thereby mitigating the spectral alignment problem.
Inspired by the auditory model CASP~\cite{jepsen2008computational}, which incorporates a modulation filterbank~\cite{dau1997modeling} to explain human sensitivity to temporal–spectral fluctuations, we propose the affine modulation-based audiogram fusion frequency–temporal Conformer (AMFT-Conformer) to effectively fuse audiogram and spectral features.
Unlike existing in-context fusion strategies that merely concatenate shallowly aligned features before encoding, our approach first applies independent encoders to transform the audiogram and spectral modalities. 
It then employs affine modulation, where the audiogram serves as a conditional input to directly scale, shift, and gate the statistical distribution of deep spectral representations, functioning as a neural analogue of auditory modulation filtering to adaptively inject individualized information across both frequency and temporal dimensions.
This design allows the model to adaptively inject individualized information across both frequency and temporal dimensions, providing a more natural way to capture how specific levels of hearing loss influence speech perception across different frequency bands and time segments.
Furthermore, since the FIG6 prescription formula tends to introduce excessive gain in non-speech segments and thereby causes distortion, we incorporate a voice activity detection (VAD) auxiliary training task. 
This enables AFN-HearNet to effectively distinguish between speech and non-speech segments, ultimately improving overall performance.
The main contributions of this work can be summarized as follows:
\begin{itemize}
\item We propose a novel audiogram-specific encoder that leverages a linear interpolation strategy to expand sparse audiogram measurements into dense representations. This interpolation-based alignment yields a physiologically plausible feature representation, enabling more effective integration with speech spectra.
\item We propose an affine modulation-based audiogram fusion frequency–temporal Conformer, which incorporates audiogram information as conditional auxiliary input to modulate noisy speech spectra, thereby achieving adaptive cross-modal fusion that injects individualized hearing profiles into spectral representations across both frequency and temporal dimensions.
\item We incorporate a VAD auxiliary task that implicitly drives the AMFT-Conformer to learn speech activity information, enabling AFN-HearNet to identify and focus on segments containing speech.
\item Comprehensive experiments are conducted across multiple datasets to validate the effectiveness of each proposed module. 
The results demonstrate that the AFN-HearNet outperforms state-of-the-art (SOTA) in-context fusion joint models in key metrics such as HASQI and PESQ.
\end{itemize} 

The remainder of the paper is organized as follows: Section~\ref{sec_related} introduces related works. Section~\ref{section_pf} presents the problem formulation, followed by a detailed description of the proposed framework in Section~\ref{section_system_desc}. The experimental setup is outlined in Section~\ref{section_experimental_setup}, while Section~\ref{section_results_and_analysis} covers the results and analysis. Finally, conclusions are presented in Section~\ref{section_conclusion}.

\section{Related work}
\label{sec_related}

\subsection{DNN-based denoising and hearing compensation methods}

\begin{figure*}[t]
	\setlength{\abovecaptionskip}{-0.2cm}
	\centering
	\includegraphics[width=\linewidth]{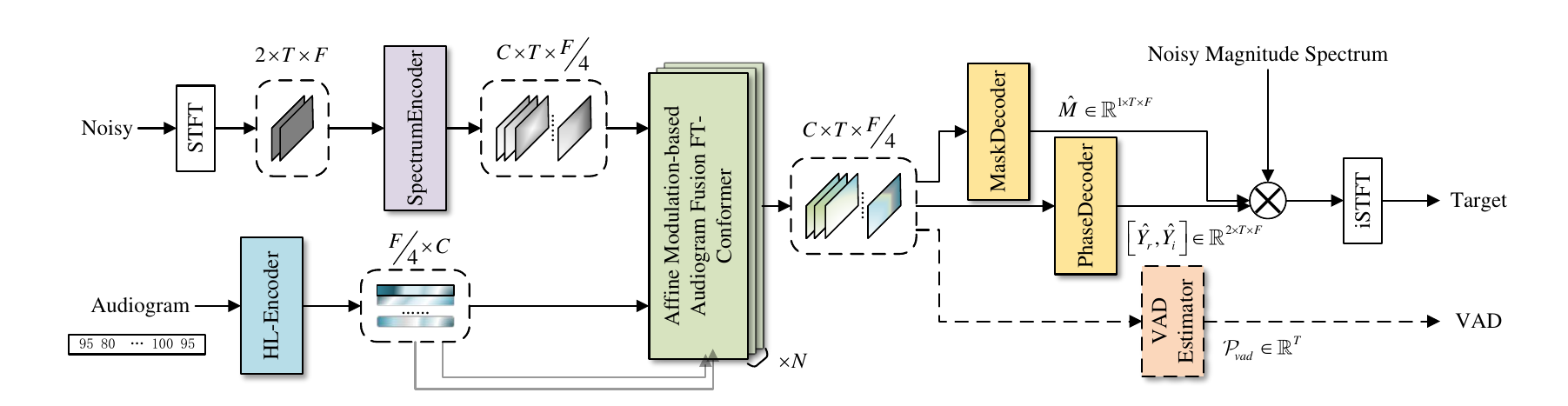}
	\caption{The overall architecture of the AFN-HearNet model.}
	\label{fig_overall_arch}
	\vspace{-0.2cm}
\end{figure*}

Due to their excellent nonlinear approximation and temporal modeling capabilities, DNN-based models~\cite{hao2023neural,zhang2023deep, mehrish2023review,xu2024adaptive,li2024dual,hu2025distributed} have shown remarkable performance in speech enhancement tasks, making them a promising solution for PSE tasks.
Most existing DNN-based approaches follow the conventional HA processing pipeline, focusing on optimizing either the NR or the HLC component independently, or sequentially combining the two.
For the NR task, prior studies~\cite{schroter2022low, tammen2022deep, lei2023low, westhausen2024real} explore the potential of the DNN-based NR components for HAs, where filter coefficients are estimated by a DNN model based on input features.
Regarding the HLC task, studies~\cite{drgas2023dynamic,leer2025hearing} propose neural networks for compensating sensorineural hearing loss.
While these approaches improve audibility, they typically optimize each component in isolation, without explicitly modeling the interactions between NR and HLC.
To address the interactions between NR and HLC, two-stage joint optimization frameworks~\cite{tu2021two,zhang2025neural} are proposed, in which both components are trained under a unified objective.
In contrast, joint models~\cite{cheng2023speech,gonzalez2025controllable} employ a unified architecture to simultaneously perform both NR and HLC tasks.
For instance, the HA-MGAN~\cite{cheng2023speech} jointly addresses NR and HLC tasks by fusing audiogram and spectral features, whereas BSRNN~\cite{gonzalez2025controllable} employs a multi-task learning framework to achieve joint optimization of the two tasks.
However, although the existing in-context fusion in joint models demonstrates effectiveness, it is still constrained by unresolved issues such as audiogram–spectrum alignment and temporal mismatch.
Current approaches typically concatenate the two modality features directly, neglecting cross-modal interactions and relying on overly simplistic fusion mechanisms. 
As a result, the subsequent encoder is forced to implicitly reconcile information from misaligned features, which constrains overall performance.

In practice, it is infeasible to measure frequency-wise paired data between audiogram information and spectral features, making precise alignment unattainable. 
To address this resolution alignment problem, HEROS-GAN~\cite{wang2025heros} leverages optimal transport theory to maximize supervisory information from unpaired data and injects Laplacian energy into the generator, transforming low-cost sensor signals into high-cost equivalents.
Previous studies~\cite{cheng2023speech,drgas2023dynamic,liang2024non,yang2025non} utilize a repeat-embedding method to expand the audiogram before concatenating it with spectral features, where each sub-band is assigned the same hearing threshold. 
In contrast, \cite{gonzalez2025controllable} employs a learnable fully connected layer to directly extract gated conditional features for deep spectral representations, thereby skipping the alignment step. 
Different from prior approaches, our method combines a psychoacoustically inspired linear interpolation strategy for audiogram expansion with a dedicated audiogram-specific encoder, enabling the extraction of discriminative patterns that more faithfully capture individualized hearing characteristics.

\subsection{Fusion schemes for integrating cross-model features}

Multi-modal fusion plays a critical role in personalized speech enhancement, where effective integration of audiogram and acoustic features is essential.
Beyond basic in-context fusion, more sophisticated mechanisms are proposed for multi-modal feature integration.
The gated fusion mechanism~\cite{fan2020gated,narayanan2020gated} adaptively regulates the information flow by learning to select and integrate the most relevant components from noisy and enhanced features.
Bilinear pooling~\cite{carreira2012semantic,lin2015bilinear,gao2016compact} employs outer product operations to capture second-order dependencies between modalities, enabling more expressive cross-feature representations.
Attention-based fusion~\cite{dai2021attentional,ni2024msa} leverages feature representations from one modality to compute attention weights for another, thereby guiding the fusion process by emphasizing more informative and semantically relevant components.
However, most of the aforementioned fusion strategies implicitly assume temporal or semantic compatibility across modalities, relying on time-varying cues to estimate fusion weights or model cross-modal interactions. 
Such assumptions are not well aligned with the inherent asymmetry between audiogram and spectrum features.
In contrast, affine transformation~\cite{yousefi2021speaker} offers a principled mechanism for cross-modal feature fusion, where one modality serves as conditional auxiliary information to guide the computation of the other. 
In parallel, auditory models~\cite{dau1997modeling} employ modulation operations to capture human sensitivity to temporal–spectral fluctuations and to simulate hearing loss in both normal-hearing and hearing-impaired listeners. 
Motivated by these insights, we introduce affine modulation into time–frequency modeling and further propose a post-gating strategy to emulate auditory masking effects, thereby enabling more effective fusion of audiogram and spectrum features.

\section{Problem description}
\label{section_pf}

Considering a hearing aid that utilizes a serial concatenation configuration as shown in Fig.\,\ref{fig_example}(a), where NR is followed by HLC, with clean speech $s(n)$ and ambient noise $e(n)$, the received microphone signal $y(n)= s(n) + e(n)$ can be expressed in the frequency domain using the short-time Fourier transform (STFT) as follows:
\begin{equation}
Y(l, k)=S(l, k) + E(l, k),
\end{equation}
where $l$ and $k$ indicate the index of the frame and frequency bin, respectively. 
In the hearing aid system, a front-end NR algorithm is typically introduced to filter out noise by applying complex ratio masking $M$ to the input spectrum:
\begin{equation}
\hat{S}(l,k) = Y(l,k)M(l,k).
\end{equation}
Next, the enhanced speech is fed into a WDRC algorithm that utilizes a FIG6 fitting formula to estimate the target SPL for each sub-band based on the input sub-band SPLs and the audiogram $HL$. 
Following this, the logarithmic gain $G(l,b)$ in decibels for loudness compensation can be expressed as:
\begin{align}
G(l,b)&=\text{SPL}_{\text{target}}(l, b)-\text{SPL}_{\text{input}}(l, b), \\
\text{SPL}_{\text{target}}(l, b) &= \text{FIG6}(\hat{S}(l); HL), 
\end{align}
where $b$ denotes the sub-band index. All frequency bins within the same sub-band share the same gain.
Consequently, the total gain for the PSE output spectrum $\hat{X}$ based on the noisy microphone spectrum is given by:
\begin{equation}
\hat{X}(l, k) = Y(l,k)\left[M\left(l,k\right)\cdot \bar{G}(l, b)\right], k\in b.
\end{equation}
where $\bar{G}(l, b)=10^{\frac{G\left(l, b\right)}{20}}$.
Therefore, to develop a model that jointly enhances and compensates for the noisy speech, the core task is to determine the joint masking parameters $M(l,k)\cdot\bar{G}(l, b)$, given the individual hearing loss audiogram.

\begin{figure}[t]
	\setlength{\abovecaptionskip}{-0.2cm}
	\centering
	\includegraphics[width=\linewidth]{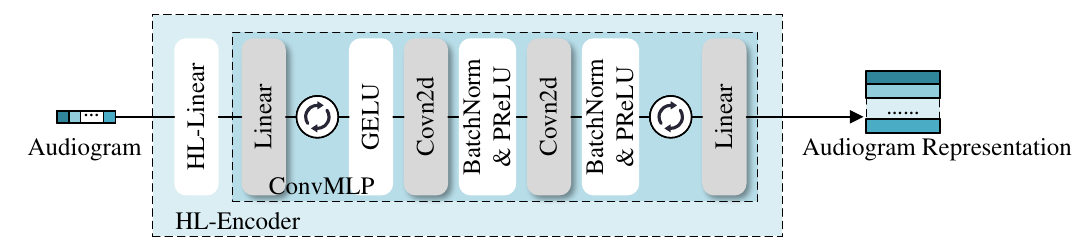}
	\caption{Detailed architecture of the audiogram encoder module.}
	\label{fig_hl_encoder}
	\vspace{-0.2cm}
\end{figure}

\section{Methodology}
\label{section_system_desc}

\subsection{Overview}

The proposed audiogram fusion PSE network can be represented as a mapping function that takes a distorted noisy waveform $Y\in\mathbb{R}^{L\times1}$ and a hearing-impaired audiogram $HL\in\mathbb{R}^{1\times N}$ as inputs and outputs the monaural de-noised and compensated waveform $\hat{X}\in\mathbb{R}^{L\times1}$, denoted as $\mathcal{F}: (Y, HL)\mapsto \hat{X}$.
The architecture of the AFN-HearNet is illustrated in Fig.\,\ref{fig_overall_arch} and consists of four key components: (1) a dual encoder structure that integrates a spectrum encoder to capture essential features while reducing dimensionality, along with an audiogram-specific HL-Encoder that transforms audiogram information for feature alignment; (2) cascaded AMFT-Conformer modules that fuses the cross-domain features by utilizing HL-Encoder outputs to adaptively modulate the unified deep representations during the frequency-wise and temporal-wise feature modeling process; (3) dual-path decoders that reconstruct the spectrum from both the magnitude and phase domains; and (4) an auxiliary VAD estimator that improves the information density and focus of unified deep representations by identifying speech-active regions.

\subsection{Spectrum encoder and decoder}

An STFT operation first converts the noisy waveform into a complex spectrogram $Y\in\mathbb{R}^{2\times T\times F}$. 
We utilize the spectrum encoder and decoders from MetricGAN~\cite{cao2022cmgan} to build the backbone network for processing and reconstructing the speech signals.
The encoder begins with two cascaded down-sample convolutional blocks, followed by a dilated DenseNet~\cite{pandey2020densely}.
Each down-sample block consists of a convolution layer, a layer normalization, and a PReLU activation function, which reduces the frequency dimension by half.
Dual-path spectrum decoders operate independently to reconstruct the output components from the fused unified deep representation produced by the cascaded AMFT-Conformer.
Each path incorporates a sub-pixel block consisting of two SPConvTransposed modules~\cite{fu2019metricgan} that up-sample the frequency dimension by a factor of two.
Specifically, the mask decoder is designed to predict a masking filter $\hat{M}\in\mathbb{R}^{T\times F}$ that will be element-wise multiplied with the input magnitude $Y_m\in\mathbb{R}^{T\times F}$.
In contrast, the phase decoder estimates the real $\hat{Y}_{r}$ and imaginary $\hat{Y}_{i}$ components of the output spectrum, allowing for the indirect reconstruction of the phase components since structural information is absent in the phase spectrogram~\cite{yin2020phasen}.
Finally, the spectrum reconstruction progress can be described as:
\begin{equation}
\begin{aligned}
\hat{X} &= (\hat{M} \odot Y_m) \cdot e^{j\angle\phi}\\
\angle\phi &= \arctan(\hat{Y}_i / \hat{Y}_r),
\end{aligned}
\end{equation}
where $\odot$ refers to the element-wise multiplication.
Finally, an inverse short-time Fourier transform is applied to convert the reconstructed spectrum back into the time domain, yielding the output compensated speech.

\subsection{Audiogram-specific encoder}

Since audiograms quantify hearing loss thresholds at a limited set of standard clinical frequencies (typically 250, 500, 1000, 2000, 4000, and 8000 Hz), their feature representation is inherently sparse. 
However, speech spectrum possess a much finer spectral resolution, and perceptually relevant spectral information may occur at intermediate frequencies not directly measured in the audiogram. 
We propose a simple yet effective method, HL-Linear, a linear interpolation technique that extends sparse clinical measurements ($\mathbb{R}^{6}\mapsto\mathbb{R}^{F}$) into a high-resolution representation aligned with the spectral resolution of the input speech.

Fig.\,\ref{fig_hl_encoder} illustrates the architecture of the audiogram-specific encoder.
Specifically, given the input audiogram $HL=[h_0,\cdots,h_5]$ measured at the frequency components $[f_0, \cdots, f_5]= [250, 500, 1000, 2000, 4000, 8000]$, the hearing loss value for any frequency component $f_k$ within the $i$-th sub-band $[f_{i+1},f_i]$ can be computed as:
\begin{align}
HL(k) = h_i + \frac{h_{i+1}-h_i}{f_{i+1}-f_i}\cdot (f_k - f_i).
\end{align}
The interpolated audiogram is then processed by a convolutional multi-layer perceptron (ConvMLP) module, which acts as a transformation function $\mathcal{F}:\mathbb{R}^{F}\mapsto\mathbb{R}^{C\times F}$.
The ConvMLP module performs channel expansion by integrating an MLP and two convolution modules, leveraging the local perceptual capabilities of convolutional layers and the global feature learning capability of the MLP.
Specifically, the extended audiogram first passes through a feature-expanding linear layer, followed by a GELU activation function.
Next, it goes through two cascaded convolution blocks, each incorporating a batch normalization (BN) and a PReLU activation layer. 
Finally, the channel dimension is adjusted using a linear layer.
The transformation progress can be expressed as:
\begin{equation}
\begin{aligned}
H^\prime_{e}&=GELU\left(\mathcal{R}\left(H_{e}\cdot W_f +b_f; \mathbb{R}^{CF}\mapsto\mathbb{R}^{C\times F}\right)\right), \\
H^{\prime}_{e}&=\delta_i\left(\mathcal{B}_i\left(Conv\left(H^\prime_{e};\theta_i\right)\right)\right),~~i\in[0,1]\\
\hat{H}_{e} &= \mathcal{R}\left(H^{\prime}_{e};\mathbb{R}^{C\times F}\mapsto\mathbb{R}^{F\times C}\right)\cdot W_c + b_c,
\end{aligned}
\end{equation}
where $H_e\in\mathbb{R}^{F}$ denotes the extended audiogram; $\cal R$ refers to the rearrange operation; $\delta_i$, $\mathcal{B}_i$, and $Conv(\cdot;\theta_i)$ denote the PReLU, BN, and convolution operations of $i$-th layer. $W$ and $b$ with subscripts $f/c$ correspond to the parameters of the first and last linear layers, respectively.

\subsection{Affine modulation-based audiogram fusion}

\begin{figure}[t]
	\setlength{\abovecaptionskip}{-0.2cm}
	\centering
	\includegraphics[width=\linewidth]{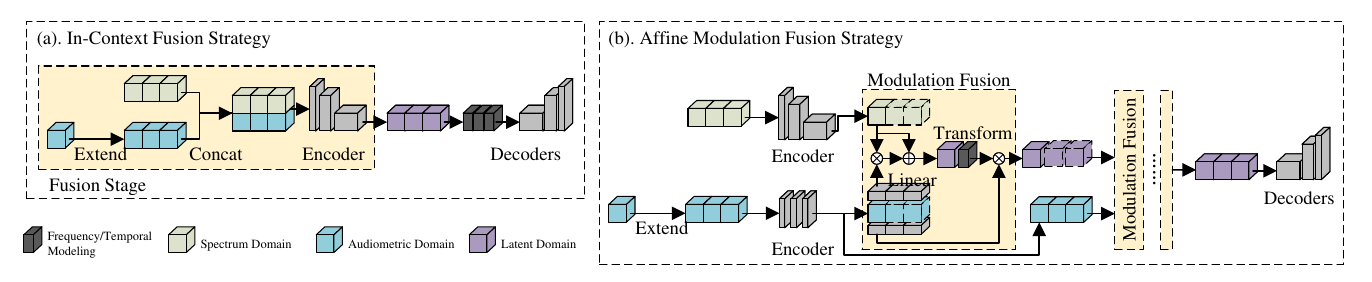}
	\caption{The differences between existing in-context and our proposed affine modulation audiogram fusion strategy.}
	\label{fig_fusion}
	\vspace{-0.2cm}
\end{figure}

\begin{figure}[t]
	\setlength{\abovecaptionskip}{-0.2cm}
	\centering
	\includegraphics[width=\linewidth]{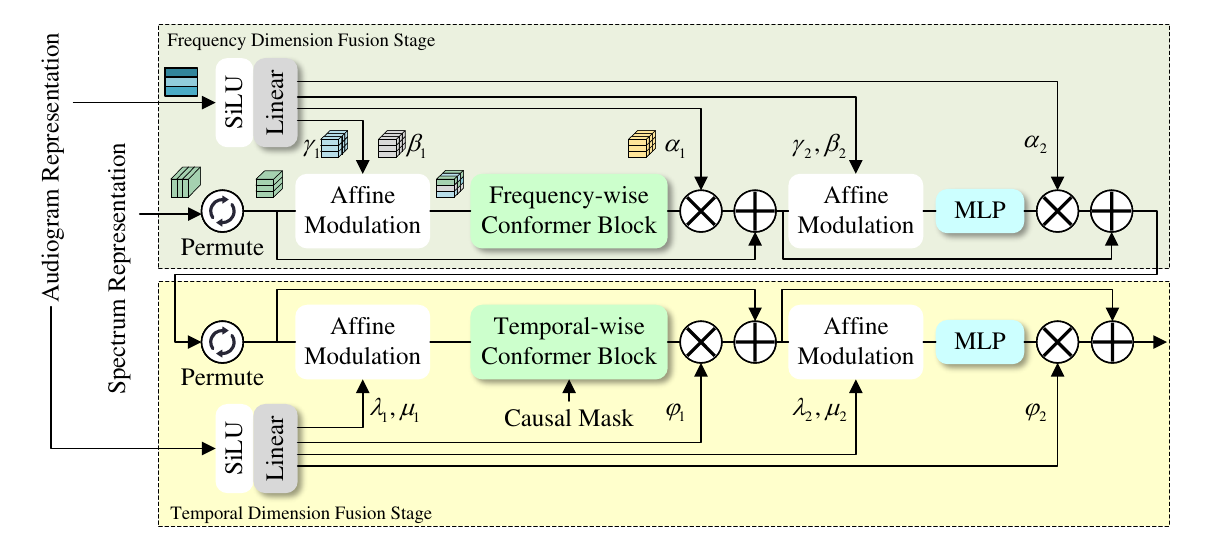}
	\caption{Details architecture of the modulation fusion AMFT-Conformer module.}
	\label{fig_amft_conformer}
	\vspace{-0.2cm}
\end{figure}

Fig.\,\ref{fig_fusion}(a) illustrates the process of the existing in-context fusion strategy.
It aligns the audiogram features with spectrum components through a feature embedding method, then concatenates these features and integrates them with a multi-layer convolutional encoder, a common fusion method for combining features from multiple models~\cite{hao2024mbfusion,hao2025mapfusion}.
However, since the fusion stage relies entirely on convolution operations and is processed only once, it fails to capture complex patterns or temporal dependencies within the features adequately, resulting in limited effectiveness.
In contrast, as shown in Fig.\,\ref{fig_fusion}(b), the affine modulation fusion strategy aligns features in the latent domain by employing independent encoding for different features. 
This fusion strategy combines affine modulation with frequency and temporal-wise Conformer modules that incorporate attention mechanisms, facilitating both local and global feature integration. 
By utilizing the cascaded structure, it effectively merges features from different levels, refining the unified deep representation to capture complex patterns.
Building on this foundation, the detailed architecture of the modulation fusion module, AMFT-Conformer, is shown in Fig.\,\ref{fig_amft_conformer}.
It consists of two fusion stages: the frequency dimension fusion stage (FFusion-Conformer) and the temporal dimension fusion stage  (TFusion-Conformer), both operating over the channel dimension.
Each stage contains a stage-oriented Conformer followed by an MLP block, with an affine modulation module positioned in front of each. 
The AMFT-Conformer takes the down-sampled deep representations from the spectrum encoder $Z\in\mathbb{R}^{C\times T\times\bar{F}}$ as the input feature, and the transformed audiogram $\hat{H}_{e}\in\mathbb{R}^{C\times\bar{F}}$ from the HL-Encoder as the personalized representation.

In the FFusion-Conformer stage, the down-sampled spectrum representation $Z\in\mathbb{R}^{C\times T\times\bar{F}}$ is first rearranged by permuting its dimensions to obtain a frequency-oriented form $Z_f\in\mathbb{R}^{T\times\bar{F}\times C}$.
The audiogram-derived representation $\hat{H}_{e}$ is processed through a SiLU–Linear transformation module, consisting of a SiLU activation followed by a linear layer, to produce two groups of frequency-adaptive affine modulation parameters, denoted as $[\gamma_i,\beta_i,\alpha_i]\in\mathbb{R}^{1\times \bar{F}\times C}$ where $i\in[1,2]$.
This progress can be represented as:
\begin{equation}
\left[\gamma_1,\beta_1,\alpha_1,\gamma_2,\beta_2,\alpha_2\right]^{T}=W_{mf} \sigma_{\text{SiLU}} (\hat{H}_{e}) + b_{mf},
\end{equation}
where $W_{mf}$ and $b_{mf}$ denote the parameters of the transformation linear layer.
These parameters enable stage-specific feature modulation, where the first group $(\gamma_1, \beta_1, \alpha_1)$ scales and shifts the frequency-oriented representation $Z_f$ prior to processing by the frequency-wise Conformer block.
In this way, the Conformer’s self-attention and convolutional submodules operate on features that have already been re-weighted and aligned to the listener’s hearing profile.
The subsequent gating operation further allows the network to selectively enhance frequency bands of higher perceptual importance while suppressing less relevant spectral components.
The frequency-wise modulation process can be expressed as:
\begin{equation}
Z^\prime_f = \left[\mathcal{T}\left(Z_f\odot\left(I+\gamma_1\right)+\beta_1\right)\right]\odot\alpha_1+Z_f,
\end{equation}
where $\mathcal{T}$ refers to the Conformer block, and $I$ denotes the all-ones matrix $\in\mathbb{R}^{1\times \bar{F}\times C}$.
Similarly, the second post-attention modulation step, parameterized by $(\gamma_2, \beta_2, \alpha_2)$, re-scales, shifts, and gates the intermediate representation, thereby introducing additional non-linear transformations that reallocate channel-wise energy while preserving the inherent sequential structure of the features.
This operation can be computed as:
\begin{align}
\hat{Z}_f &= \left[MLP\left(Z^\prime_f\odot\left(I+\gamma_2\right)+\beta_2\right)\right]\odot\alpha_2+Z^\prime_f.
\end{align}

The TFusion-Conformer stage adopts the same architecture as the FFusion-Conformer but operates along the temporal dimension, applying modulation and Conformer processing to time-oriented feature sequences in the form $Z_t\in\mathbb{R}^{\bar{F}\times T\times C}$.
While the audiogram is inherently time-invariant, effective compensation is time-dependent, as the compensation gain must be adapted to the dynamic characteristics of speech.
Temporal-wise modeling enables the network to capture such variations, including transient events and fluctuations in speech energy. 
In the AMFT-Conformer, this capability is realized through the TFusion-Conformer stage, which performs temporal fusion to integrate personalized hearing profiles with dynamic speech features for time-aware compensation.

\subsection{VAD auxiliary module}

Since the FIG6 compensation algorithm estimates the target gain based on the input SPL, non-speech segments become more susceptible to the influence of residual noise. 
In the case of high SPL background noise, the preceding speech enhancement algorithm has already suppressed its stationary components, thereby substantially reducing the residual noise energy and bringing the SPL closer to that of quiet segments.
Consequently, the SPL of non-speech segments is reduced to extremely low levels. 
Although these noise components are of low energy, they still fall within the FIG6 amplification region, which is typically inactive only below 20\,dB SPL, a condition that is rarely met in real-world acoustic environments.
As a result, these components are subject to excessive gain amplification, leading to perceptible distortion in non-speech segments, as illustrated in Fig.\,\ref{fig_example_2}.

We propose a VAD auxiliary training module that takes the unified deep representation output by the AMFT-Conformer as input. 
This module implicitly encodes latent feature patterns of both speech and non-speech segments within the unified deep representation, effectively building awareness of the acoustic context in the feature latent space. 
This information about the acoustic context is particularly useful for addressing the noise amplification issue in silent segments caused by the WDRC loudness compensation algorithm based on the FIG6 formula while maintaining the model's inference efficiency.
Since the VAD estimator is only active during training, it introduces no additional parameters or computational overhead during deployment.

The module begins with an average pooling layer to aggregate the global features of each channel, followed by a point-wise convolution block, where each convolution operation is succeeded by layer normalization and a PReLU activation function. 
Next, we employ a GRU layer to model temporal dependencies, effectively capturing sequential patterns and variations over frames. 
Finally, the output is fed into a fully connected layer that reduces the dimensionality and passes through a sigmoid function to estimate the speech activity probabilities.
Given the input unified deep representation $H\in\mathbb{R}^{C\times T\times F}$, the process of estimating the VAD probability $P_{vad}\in\mathbb{R}^{T\times 1}$ can be expressed as follows:
\begin{equation}
\begin{aligned}
P_{vad}&=\sigma\left(\text{GRU}\left(H^\prime\right)\cdot W+b\right),\\
H^\prime&=\delta\left\{\mathcal{B}_{LN}\left[\mathcal{R}\left(\text{PWConv}\left({\cal G}\left(H\right)\right); 
\mathbb{R}^{C\times T\times 1}\mapsto\mathbb{R}^{T\times C}\right)\right]\right\},
\end{aligned}
\end{equation}
where 
$\mathcal{G}(H)=\frac{1}{F}\sum\nolimits_{i=0}^{F-1}H[..., i]\in\mathbb{R}^{C\times T\times 1}$ represents the average pool operation across the frequency dimension for each channel, and $\text{PWConv}$ refers to the convolution layer with $(1,1)$ kernels.
Symbols $\sigma$, $\delta$, $\mathcal{B}_{LN}$, and $\cal{R}$ denote the sigmoid, PReLU, layer normalization, and rearrange operations, respectively.
$W$ and $b$ represent the parameter matrices of the linear layer.

\subsection{Loss function}

In this work, we introduce the Hearing Aid Speech Quality Index (HASQI) metric~\cite{kressner2012evaluating} as a key optimization target due to its established role in evaluating HLC effectiveness. 
Since the HASQI function is non-differentiable, we adopt the MetricGAN framework~\cite{fu2019metricgan,cheng2023speech}, which enables gradient-based optimization of black-box metrics through adversarial training.
For spectrum reconstruction objectives, we introduce the multi-resolution STFT loss~\cite{defossez2020real}, which operates on the spectrogram magnitudes.
This loss is an expectation of various STFT losses configured with different frames and hop lengths, which can be computed as:
\begin{equation}
\begin{aligned}
\mathcal{L}_{stft}(X, \hat{X}) &= 0.5 \times(\mathcal{L}_{sc}(X, \hat{X}) + \mathcal{L}_{mag}(X, \hat{X})),\\
\mathcal{L}_{sc}(X, \hat{X}) &=\frac{\Vert~|\text{STFT}({X})|-|\text{STFT}(\hat{X})|~\Vert_{F}}  {\Vert~|\text{STFT}({X})|~\Vert_{F}},\\
\mathcal{L}_{mag}(X, \hat{X}) &= \frac{1}{T} \Vert \log|\text{STFT}(X)| - \log|\text{STFT}(\hat{X})|~\Vert_{1},
\end{aligned}
\end{equation}
where the $\Vert \cdot \Vert_F$ and $\Vert \cdot \Vert_1$ represent the Frobenius and $L_1$ norms, respectively. 
$X$ denotes the compensated clean target and $\hat{X}$ refers to the enhanced speech of the AFN-HearNet model. 
In this work, we setup the frame sizes to $[1024, 512, 256]$ and corresponding hop lengths to $[512, 256, 128]$.
Additionally, we introduce the perceptual metric for speech quality evaluation (PMSQE)~\cite{martin2018deep} loss to optimize for human auditory perception, and employ a focal loss to train the VAD auxiliary task.

Overall, given the hearing-impaired audiogram $HL$, compensated clean speech $X$, and enhanced output $\hat X$, and the ground-truth as well as estimated VAD labels $V$ and $\hat{V}$, the joint losses for the entire framework are described as follows:
\begin{equation}
\label{loss_eqa}
\begin{aligned}
{\cal L}_G &=\alpha\cdot \mathbb{E}\left[ \left\| D_{\text{fix}}\left( X,\hat X,HL \right) - 1 \right\|^2 \right]\\
&+ \lambda\cdot{\cal L}_{PMSQE}\left( X,\hat X \right)
+ {\cal L}_{stft}\left(X,\hat X\right)+0.3\cdot\mathcal{L}_{focal}(V,\hat{V}),\\
{\cal L}_D &= \mathbb{E}\left[\left\| {D\left(X,X,HL\right)-1} \right\|^2\right] \\ 
&+ \mathbb{E}\left[\left\| D\left(X,\hat X,HL\right)-\text{HASQI}\left(X,\hat X,HL\right) \right\|^2 \right],
\end{aligned}
\end{equation}
where ${\cal L}_G$ and ${\cal L}_D$ denote the loss functions to train the generator of AFN-HearNet and a corresponding discriminator of $D$ proposed in MetricGAN, respectively. 
$\alpha$ and $\lambda$ are used to balance the various loss values.
Given that GAN training is highly sensitive to the weighting of loss terms~\cite{pan2020loss}, a further analysis is presented in Appendix~\ref{appdix_a}.

\section{Experiments}
\label{section_experimental_setup}

\subsection{Datasets}

\begin{algorithm}[t]
	\footnotesize
	\caption{The procedure of synthesizing samples.}
	\label{algo_synth_dset}
	\SetAlgoLined
	\KwIn{The total number of synthetic samples $N$, and the duration of each sample $T$.}
	\KwData{Audio dataset $\mathcal{D}$, and audiogram collection $\mathcal{H}$.}
	\KwOut{Loudness compensation dataset $\mathcal{D}_{train}$.}
	
	\For{$N$ iterations}{
		\tcp{1. Prepare the speech and the noisy pair.}
		Initialize collection $\mathcal{D}_{train}=[]$\;
		\Repeat{$\text{activity}(s)\ge 0.6$}{
			Sample a clean speech $s_0$ from $\mathcal{D}$\;
			$t \gets \mathcal{U}(0, \text{len}(s)-T)$\;
			$s \gets s_0[t:t+T]$\;
		}
		Sample a noise $e$ signal from $\mathcal{D}$\;
		Sample mode from ['release', 'attack', 'bypass'] with $p=[0.4,0.3, 0.3]$\;
		\uIf{'release' == mode}{ \tcc{sample a gain from uniform distribution to amplify speech}
			Sample a gain $G$ from $\mathcal{U}(\text{rms}_{dB}(s)+5,-10)$\;
		}\uElseIf{'attack' == mode}{ \tcc{gain to attenuate speech}
			Sample a gain $G$ from $\mathcal{U}(-35,\text{rms}_{dB}(s)-5)$\;
		}
		$x_{G}\gets \text{applyGain}(s,G,t_d)$\;
		$e\gets + $ adding gaussian noise in range $[\text{rms}_{dB}(e)-10, \text{rms}_{dB}(e)]$ with $p=0.7$\;
		Sample a SNR from $[-5, 15]$\;
		$e_{snr}\gets$ Adjust noise $e$ based on speech $x_{G}$ and SNR\;
		$x_{noisy}\gets s_{G}+e_{snr}$\;
		\tcp{2. Hearing compensation.}
		Sample an audiogram $hl$ from $\mathcal{H}$\;
		\tcp{Compensation with VAD information.}
		$s_{hl}\gets \mathcal{F}_{vad}(\text{WDRC}(s_{G}, hl),s_G)$\;
		\KwResult{One sample ($x_{noisy}, hl$) with target $s_{hl}$\;}
		$\mathcal{D}_{train}\gets (x_{noisy}, hl,s_{hl})$\;
	}
\end{algorithm}

To conduct the experiments, we utilized two synthetic datasets derived from widely used public datasets: one dataset based on the DNS-Challenge~\cite{reddy2020interspeech} and another based on the LibriSpeech~\cite{panayotov2015librispeech}+Demand~\cite{thiemann2013demand} dataset.
The DNS-Challenge\footnote{https://github.com/microsoft/DNS-Challenge} dataset consists of 500\,h clean speech from 2150 speakers and 180\,h noise.
The LibriSpeech\footnote{https://www.openslr.org/12/} corpus consists of approximately 1000\,h of read English speech, and we utilize the clean speech from several subsets, i.e., train-clean-100, dev-clean, and test-clean,
along with noise from the Demand\footnote{https://zenodo.org/records/1227121\#.YXtsyPlBxjU} corpus.
The audiograms used in the experiments were sourced from the National Health and Nutrition Examination Survey (NHANES)~\cite{cheng2023speech}. 
The dataset consists of 114 audiograms obtained following standard measurement procedures for pure-tone audiometry, covering ascending, sloping, and flat audiogram curve types.
Each participant contributed both left- and right-ear audiograms. 
The average age of the hearing-impaired participants was 68.4 years, with 27 males and 30 females.
The mean hearing threshold at 500\,Hz, 1\,kHz, 2\,kHz, and 4\,kHz is used to calculate pure tone average, and the degree of deafness was categorized according to the severity of the hearing loss: 10 audiograms with thresholds below 40\,dB were classified as mild or less severe; 61 audiograms with thresholds between 40\,dB and 70\,dB as moderate or moderately severe; and 43 audiograms exceeding 70\,dB as severe or profound~\cite{alshuaib2015classification}. 

The detailed procedures for synthesizing datasets are outlined in Algorithm~\ref{algo_synth_dset}. 
First, sample a clean speech $s$ that exhibits over 60\,\% speech activity, along with a noise $e$.
To simulate different scenarios of loudness compensation, where small signals are released while large signals are attenuated according to their SPL, we transform the clean speech $s$ by either amplifying, suppressing, or leaving unchanged, with respective probabilities of $[0.4, 0.3, 0.3]$.
The transformed clean speech is denoted as $x_G$.
During the amplification phase, the clean speech energy is increased by at least 5\,dB, but not exceeding an upper limit of -10\,dB.
Conversely, in the suppression phase, the clean speech energy decreases by at least 5\,dB, reaching a minimum of $-35$\,dB.
After that, with a probability of 0.7, Gaussian white noise is added to the sampled noise. 
The noisy $x_{noisy}$ is synthesized by combining the clean speech with this noise according to an SNR sampled from a uniform distribution ranging from $[-5,15]$\,dB. 
Then, we apply the WDRC algorithm, based on the FIG6 fitting formula, which takes the transformed speech $s_G$ and audiogram vector $hl$ as inputs, to generate the target speech $s_{hl}$. 
To address the problem of excessive gain in silent segments resulting from the FIG6 fitting formula, we replace the compensated non-speech segments of target speech with the original speech segments before compensation.
In total, we synthesized 72000 training samples ($\sim$100\,h) and 1440 validation samples ($\sim$2\,h), each lasting 5\,s, for both the DNS-Challenge dataset and LibriSpeech+Demand dataset.
The no-reverberation test set of the DNS-Challenge dataset contains 150 samples, while the LibriSpeech+Demand test set includes 1440 samples ($\sim$2\,h) for comparison with other baselines.
Ablation studies are conducted on the DNS-Challenge dataset, with evaluations performed on both the no-reverberation DNS-Challenge test set and a synthesized test.
This synthesized test set combines speech from the LibriSpeech test-clean subset with noise from the NoiseX-92 corpus~\cite{varga1993ii} under SNRs from $-5$\,dB to 15\,dB in 5\,dB increments, with each SNR containing 720 samples ($\sim$1\,h).
The NoiseX-92 corpus consists of 15 different types of real-world noise recordings, including babble, factory noise, and street noise.

\begin{table}[t]
	\caption{The proposed model architecture. $T$ denotes the number of frames.}
	\label{table_model_arch}
	\centering
		\begin{tabular}{l*{3}{>{\centering\arraybackslash}p{2.7cm}}}
			\toprule
			Module & Input size & Hyperparameters & Output size \\
			\midrule
			Spectrum Encoder & $2\times T\times 257$ & - & $48\times T\times 65$ \\
			~~~~- Down-sample\_1 & $2\times T\times 257$ & $1\times 3,(1,2)$ & $48\times T\times 129$  \\
			~~~~- Down-sample\_2 & $48\times T\times 129$ & $1\times 3,(1,2)$ & $48\times T\times 65$  \\
			~~~~- DilatedDense\_i & $48\times T\times 65$ & $2\times 3,(1,1),(2^i,1)$ & $48\times T\times 65$  \\
			Audiogram Encoder & $1\times 6$ & - & $65\times 48$ \\
			~~~~- HL-Linear & $1\times 6$ & - & $1\times 257$  \\
			~~~~- Linear\_1 & $1\times 257$ & $(257,1028)$ & $1\times 1028$  \\
			~~~~- Permute\_1 & $1\times 1028$ & - & $4\times 1\times 257$  \\
			~~~~- Conv2d\_1 & $4\times 1\times 257$ & $ 1\times 3, (1,2)$ & $16\times 1\times 129$  \\
			~~~~- Conv2d\_2 & $16\times 1\times 257$ & $ 1\times 3, (1,2)$ & $64\times 1\times 65$  \\
			~~~~- Permute\_2 & $64\times 1\times 65$ & - & $65\times 64$  \\
			~~~~- Linear\_2 & $65\times 64$& $(64, 48)$ & $65\times 48$  \\
			\midrule
			AMFT-Conformer & \makecell{$48\times T\times 65$\\ $65\times 1\times 48$} & - & $48\times T\times 65$ \\
			~~~~- F-Conformer & $T\times 65\times 48$ & Heads: 4 & $T\times 65\times 48$ \\
			~~~~- T-Conformer & $65\times T\times 48$ & Heads: 4 & $65\times T\times 48$ \\
			\midrule
			Mask Decoder & $48\times T\times 65$ & - & $1\times T\times 257$ \\
			~~~~- Sub-pixel\_1 & $48\times T\times 65$ & \makecell{$1\times3,(1,1)$ \\ $1\times2,(1,1)$} & $48\times T\times 129$ \\
			~~~~- Sub-pixel\_2 & $48\times T\times 129$ & \makecell{$1\times3,(1,1)$ \\ $1\times2,(1,1)$} & $1\times T\times 257$ \\
			Phase Decoder & $48\times T\times 65$ & - & $2\times T\times 257$ \\
			~~~~- Sub-pixel\_1 & $48\times T\times 65$ & \makecell{$1\times3,(1,1)$ \\ $1\times2,(1,1)$} & $48\times T\times 129$ \\
			~~~~- Sub-pixel\_2 & $48\times T\times 129$ & \makecell{$1\times3,(1,1)$ \\ $1\times2,(1,1)$} & $2\times T\times 257$ \\
			VAD Estimator & $48\times T\times 65$ & - & $T\times 1$ \\
			~~~~- AveragePool & $48\times T\times 65$ & $1\times 65, (1,65)$ & $48\times T\times 1$  \\
			~~~~- Permute & $48\times T\times 1$ & - & $T\times 48$  \\
			~~~~- GRU & $T\times 48$ & 128 & $T\times 128$ \\
			~~~~- Linear & $T\times 128$ & $(128,1)$ & $T\times 1$  \\
			\bottomrule
		\end{tabular}
	\vspace{-0.2cm}
\end{table}

\subsection{Implementation details}

All speech recordings were sampled at a frequency of 16\,kHz.
The audio waveform is converted to the frequency domain using STFT with a Hanning window of 512 (32\,ms) and a hop size of 256 (16\,ms).
In our work, we set the number of middle channels to 48 and the number of cascaded AMFT-Conformer modules to 2.
A more detailed description of the network parameters is given in Table \ref{table_model_arch}.
The hyperparameters of convolution layers for the encoder and decoder are specified as follows: filterHeight $\times$ filterWidth, stride (along frame, frequency), and dilation (along frame, frequency) if applicable.
The parameters $\alpha, \lambda$, and $\mu$ for the discriminator, PMSQE, and multi-resolution STFT loss are set to 0.5, 0.3, and 1, respectively.
We utilize the Adam optimizer with a learning rate of $5e^{-4}$, which is decayed by a factor of 0.5 every 10 epochs.
The proposed model is trained for a total of 25 epochs.

\subsection{Comparative baselines}

To assess the performance of both benchmarks fully, we conducted a fair comparison with several baselines that jointly address the NR and HLC tasks using the in-context fusion strategy. 
This comparison includes classical models such as  CMGAN, as well as SOTA models like SEMamba and MP-SENet.

For the DNS-Challenge benchmark, we consider the following seven baselines:
WDRC-FIG6~\cite{killion19933}, a traditional HLC algorithm that employs the FIG6 fitting formula;
CRN~\cite{tan2018convolutional}, the first model developed using the encoder-decoder architecture in the field of speech enhancement;
DCCRN~\cite{hu2020dccrn}, which revises both convolutional and recurrent structures to effectively handle complex-valued operations;
CMGAN~\cite{cao2022cmgan}, a Conformer-based metric generative adversarial network that employs two-stage Conformer blocks to model temporal and frequency dependencies;
NUNet\_TLS~\cite{hwang2022monoaural} that utilizes two-level skip connections in a U-Net architecture, along with causal time-frequency attention, to improve the dynamic representation of speech context across multiple scales;
CompNet~\cite{fan2023compnet}, a dual-stage model that enhances noisy speech in the time domain and uses a post-processing module to filter magnitude and refine phase;
HA-MGAN~\cite{cheng2023speech}, the first joint framework that applies a metric generative adversarial strategy for simultaneous NR and HLC tasks.
For the LibriSpeech+Demand benchmark, we introduce the five additional baselines.
DB-AIAT~\cite{yu2022dual} is a dual-branch attention-in-attention transformer-based model developed to handle both coarse- and fine-grained regions of the spectrum in parallel.
MP-SENet~\cite{lu2023mp} concurrently denoises magnitude and phase spectra through a codec architecture with convolution-augmented transformers, and utilizes multi-level losses for joint training.
FSPEN~\cite{yang2024fspen} is an ultra-lightweight network designed for real-time speech enhancement in resource-constrained environments, utilizing both full-band and sub-band structures to extract global and local features.
SEMamba~\cite{chao2024investigation} is a scalable state-space model developed for speech enhancement, utilizing a Mamba-based regression approach to characterize speech signals while incorporating signal-level distances and metric-oriented loss functions.
PrimeK-Net~\cite{lin2025primek} is a speech enhancement framework centered on computational efficiency, and it incorporates a group prime kernel feedforward network that extracts deep temporal and frequency features with linear complexity.

\subsection{Evaluation metrics}

We employed a comprehensive set of objective metrics to evaluate both individual module contributions and overall system performance. 
For speech enhancement quality assessment, we adopt the Perceptual Evaluation of Speech Quality (PESQ)~\cite{rix2001perceptual} in both narrow and wide bands configurations, along with the Short-Time Objective Intelligibility (STOI)~\cite{taal2011algorithm} metric for quantifying speech intelligibility improvements.
To assess speech quality for hearing-impaired listeners, we incorporate the HASQI~\cite{kressner2012evaluating} metric, a key indicator of hearing aid performance.
Furthermore, we introduce the signal-to-distortion ratio (SDR)~\cite{le2019sdr}, which measures the overall distortions of the enhanced signal relative to the clean reference, and scale-invariant signal-to-noise ratio (SI-SNR)~\cite{vincent2006performance}, which provides a robust assessment of amplitude scaling.

\section{Results and analysis}
\label{section_results_and_analysis}

The experimental results are analyzed and discussed in detail in this section. 

\subsection{Ablation Study}

\begin{table*}[h]
	\caption{Ablation studies on the DNS-Challenge dataset. The best performances of each metric for the two FT-Sequence base models are highlighted in bold.}
	\label{table_ablation_study_dns}
	\centering
	\resizebox{\textwidth}{!}{%
		\begin{tabular}{l>{\centering\arraybackslash}p{1.0cm}>{\centering\arraybackslash}p{1.0cm}*{12}{>{\centering\arraybackslash}p{1.2cm}}}
			\toprule
			\multirow{2.5}{*}{Models} & 
			\#Param& FLOPs&
			\multicolumn{6}{l}{Validation set} & 
			\multicolumn{6}{l}{Test set with no reverberation} \\
			\cmidrule(lr){4-9} \cmidrule(l){10-15}
			&(M)&(G/s)& HASQI & WB-PESQ & NB-PESQ & SDR & SI-SNR & STOI (\%) & HASQI& WB-PESQ & NB-PESQ & SDR & SI-SNR & STOI (\%) \\ 
			\midrule
			Noisy &-&-&0.59&1.11&1.29&3.65&2.45&80.68 & 0.61 & 1.09 & 1.29 & 3.74 & 2.87 & 84.77   \\ %
			\midrule
			\multicolumn{15}{l}{\bf FT-LSTM based model:} \\
			HA-MGAN&0.71&2.66 &0.66&2.43&3.01&10.11&9.08&89.84&0.69&2.40&2.99&10.26&9.56&92.57  \\
			\ +VAD &0.92&2.67&0.73&2.57&3.13&12.18&11.48&91.38&0.77&2.56&3.11&12.36&11.95&94.35  \\
			\ +HL-Linear &0.71&2.66&0.72&2.57&3.13&11.78&11.00&91.20&0.76&2.57&3.13&12.01&11.54&94.16  \\
			\ +AMFT-LSTM \& HL-Encoder &1.64&3.75&0.74&2.59&3.15&12.23&11.52&91.43&0.77&\bf2.64&\bf3.19&12.42&11.98&94.46  \\
			\ +AMFT-LSTM \& HL-Encoder \& VAD  &1.86&3.76&\bf0.74&\bf2.59&\bf3.15&\bf12.40&\bf11.71&\bf91.58&\bf0.78&2.62&3.17&\bf12.59&\bf12.14&\bf94.55  \\
			\midrule\midrule
			\multicolumn{15}{l}{\bf FT-Conformer based model:} \\
			AFA-HearNet-base &0.71&3.07&0.76&2.60&3.16&13.42&12.95&91.94& 0.80 &2.70&3.23&13.57&13.33& 95.17  \\
			\ +VAD &0.88&3.08&0.76&2.59&3.15&13.48&13.00&92.10& 0.80&2.68&3.21&13.63&13.39& 95.31 \\
			\ +HL-Linear &0.71&3.07&0.76&2.61&3.17&13.58&13.11&92.19& 0.80&2.71&3.24&13.72&13.48& 95.34 \\
			\ +AMFT-Conformer \& HL-Encoder &1.11&3.38&0.77&2.65&3.20&\bf 13.86&\bf 13.39&92.51& 0.81&2.75&3.27&\bf 13.92&\bf 13.68& 95.51 \\
			\ +AMFT-Conformer \& HL-Encoder \& VAD (prop.)  &1.28&3.39&\bf 0.77&\bf 2.67&\bf 3.22&13.75&13.28&\bf 92.57&\bf 0.81&\bf 2.75&\bf 3.29&13.86&13.62&\bf  95.55 \\
			\bottomrule
		\end{tabular}%
	}
	\vspace{-10pt}
\end{table*}

\begin{table*}[h]
	\caption{Ablation studies on the LibriSpeech+NoiseX-92 test set, where speech and noise were mixed at various SNR levels.
	}
	\label{table_ablation_study_libri}
	\centering
	\renewcommand{\arraystretch}{1}
	\resizebox{\textwidth}{!}{%
		\begin{tabular}{lcccccccccccccccccc}
			\toprule
			\multirow{3.5}{*}{Models} & 
			\multicolumn{6}{l}{HASQI} & 
			\multicolumn{6}{l}{WB-PESQ} &
			\multicolumn{6}{l}{NB-PESQ} \\
			\cmidrule(lr){2-7} \cmidrule(lr){8-13} \cmidrule(l){14-19}
			& SNR &&&&&&SNR&&&&&&SNR&&&&& \\
			\cmidrule(l){2-19} 
			& -5\,dB & 0\,dB & 5\,dB & 10\,dB & 15\,dB & Avg & -5\,dB & 0\,dB & 5\,dB & 10\,dB & 15\,dB & Avg & -5\,dB & 0\,dB & 5\,dB & 10\,dB & 15\,dB & Avg \\ 
			\midrule
			Noisy &0.15&0.21&0.29&0.33&0.38&0.27&1.04&1.04&1.06&1.07&1.11&1.06&1.13&1.17&1.22&1.29&1.38&1.24   \\ %
			\midrule
			\multicolumn{19}{l}{\bf FT-LSTM based model:} \\
			HA-MGAN &0.40&0.51&0.60&0.65&0.70&0.57 &1.32&1.57&1.90&2.20&2.52&1.90&1.80&2.20&2.64&2.97&3.27&2.58  \\
			\ +VAD &0.43&0.56&0.67&0.72&0.78&0.63
			&1.36&1.62&1.99&2.32&2.70&2.00
			&1.83&2.26&2.73&3.10&3.43&2.67  \\
			\ +HL-Linear &0.43&0.55&0.66&0.72&0.77&0.63 &1.36&1.63&2.01&2.34&2.72&2.01&1.85&2.29&2.76&3.12&3.44&2.69 \\
			\ +AMFT-LSTM \& HL-Encoder  &0.44&0.56&0.67&0.73&0.78& 0.64
			&1.38&1.65&\bf2.03&2.37&\bf2.75&\bf2.04
			&1.87&2.31&2.78&3.14&3.47&2.71  \\
			\ +AMFT-LSTM \& HL-Encoder \& VAD &\bf0.45&\bf0.57&\bf0.68&\bf0.73&\bf0.79&\bf0.64
			&\bf1.38&\bf1.65&2.02&\bf2.38&2.74&2.03
			&\bf1.87&\bf2.31&\bf2.78&\bf3.15&\bf3.47&\bf2.72  \\
			\midrule\midrule
			\multicolumn{15}{l}{\bf FT-Conformer based model:} \\
			AFA-HearNet-base&0.45&0.58&0.69&0.75&0.81&0.66
			&1.40&1.67&2.06&2.43&2.83&2.08
			&1.87&2.32&2.80&3.19&3.54&2.74  \\
			\ +VAD &0.45&0.58&0.69&0.76&0.81&0.66
			&1.39&1.66&2.03&2.40&2.79&2.05
			&1.86&2.29&2.77&3.17&3.52&2.72  \\
			\ +HL-Liner &0.45&0.58&0.70&0.76&0.82&0.66
			&1.40&1.68&2.06&2.44&2.83&2.08
			&1.88&2.32&2.81&3.20&3.54&2.75  \\
			\ +AMFT-Conformer \& HL-Encoder  &0.47&0.59&0.70&0.76&0.82&0.67
			&1.42&1.70&2.08&2.44&2.83&2.09
			&1.91&2.35&2.83&3.21&3.55&2.77  \\
			\ +AMFT-Conformer \& HL-Encoder \& VAD (prop.)   &\bf0.47&\bf0.59&\bf0.70&\bf0.76&\bf0.82&\bf0.67
			&\bf1.44&\bf1.71&\bf2.09&\bf2.45&\bf2.84&\bf2.11
			&\bf1.93&\bf2.37&\bf2.85&\bf3.23&\bf3.56&\bf2.79  \\
			\midrule
			\multirow{3.5}{*}{Models} & 
			\multicolumn{6}{l}{SDR} & 
			\multicolumn{6}{l}{SI-SNR} &
			\multicolumn{6}{l}{STOI\,(\%)} \\
			\cmidrule(lr){2-7} \cmidrule(lr){8-13} \cmidrule(l){14-19}
			& SNR &&&&&&SNR&&&&&&SNR&&&&& \\
			\cmidrule(l){2-19} 
			& -5\,dB & 0\,dB & 5\,dB & 10\,dB & 15\,dB & Avg & -5\,dB & 0\,dB & 5\,dB & 10\,dB & 15\,dB & Avg & -5\,dB & 0\,dB & 5\,dB & 10\,dB & 15\,dB & Avg \\ 
			\midrule
			Noisy &-6.03&-1.62&1.20&3.22&4.36&0.23&-6.58&-2.31&1.20&3.22&4.36&-0.02&67.36&73.20&78.84&81.87&84.98&77.25   \\ %
			\midrule
			\multicolumn{15}{l}{\bf FT-LSTM based model:} \\
			HA-MGAN &3.23&6.25&8.47&9.87&11.04&7.77&2.21&5.34&7.63&8.91&10.05&6.83 &75.48&83.02&88.09&90.14&92.41&85.83  \\
			\ +VAD &3.45&6.90&9.72&11.75&13.55&9.07
			&2.61&6.21&9.15&11.09&12.89&8.39
			&76.39&84.24&89.50&91.73&94.10&87.19  \\
			\ +HL-Linear &3.54&6.87&9.59&11.47&13.10&8.91&2.68&6.16&8.97&10.75&12.37&8.19 &76.73&84.35&89.51&91.66&93.96&87.24  \\
			\ +AMFT-LSTM \& HL-Encoder &\bf3.72&\bf7.09&9.85&11.84&13.65&9.23
			&\bf2.84&\bf6.37&9.25&11.16&12.98&8.52
			&76.67&84.54&89.69&91.83&94.13&87.37  \\
			\ +AMFT-LSTM \& HL-Encoder \& VAD &3.64&7.07&\bf9.92&\bf12.00&\bf13.86&\bf9.30
			&2.78&6.35&\bf9.31&\bf11.30&\bf13.18&\bf8.58
			&\bf77.19&\bf84.76&\bf89.83&\bf91.94&\bf94.24&\bf87.59  \\
			\midrule\midrule
			\multicolumn{15}{l}{\bf FT-Conformer based model:} \\
			AFA-HearNet-base&3.77&7.36&10.42&12.76&14.99&9.86
			&2.91&6.71&9.95&12.25&14.55&9.27
			&77.43&85.25&90.45&92.52&94.77&88.08  \\
			\ +VAD  &3.74&7.31&10.36&12.77&15.08&9.85
			&2.91&6.69&9.92&12.28&14.65&9.29
			&77.42&85.27&90.50&92.62&94.90&88.14  \\
			\ +HL-Linear &3.97&7.49&10.50&12.87&15.15&10.00
			&3.08&6.84&10.03&12.38&14.73&9.41
			&77.35&85.40&90.56&92.67&94.91&88.18  \\
			\ +AMFT-Conformer \& HL-Encoder  &4.06&7.61&10.67&\bf13.02&\bf15.33&10.14
			&3.23&6.94&\bf10.19&\bf12.51&\bf14.89&\bf9.55
			&\bf78.60&85.90&90.81&92.83&95.04&\bf88.64  \\
			\ +AMFT-Conformer \& HL-Encoder \& VAD (prop.)   &\bf4.15&\bf7.66&\bf10.67&12.97&15.26&\bf10.14
			&\bf3.25&\bf6.98&10.17&12.46&14.82&9.54
			&78.39&\bf85.92&\bf90.84&\bf92.89&\bf95.07&88.62  \\
			\bottomrule
		\end{tabular}%
	}
	\vspace{-10pt}
\end{table*}

\begin{table}[h]
	\caption{The Student's t-test $p$-value for the significance analysis of the improvements of the proposed modules compared to the baseline model, where $\ssymbol{2}$ and $\ssymbol{3}$ denote the significance analyses for the FT-LSTM and FT-Conformer base model, respectively. Superscript $^+$ indicates that the improvement of two pairs is statistically significant at the 95\% confidence level. DNS and Libri refer to the DNS-challenge and LibriSpeech+NoiseX-92 test sets, respectively.}
	\label{table_significance}
	\centering
	\resizebox{\linewidth}{!}{%
		\begin{tabular}{l|>{\centering\arraybackslash}p{1.3cm}|*{6}{>{\centering\arraybackslash}p{1.3cm}}}
			\toprule
			\multirow{2.5}{*}{Modules} & 
			\multirow{2.5}{*}{Test set} &
			\multicolumn{2}{c}{HASQI} & 
			\multicolumn{2}{c}{WB-PESQ} &
			\multicolumn{2}{c}{NB-PESQ} \\
			\cmidrule(lr){3-4} \cmidrule(lr){5-6} \cmidrule(l){7-8}
			&& $p$-value$\ssymbol{2}$ &$p$-value$\ssymbol{3}$ &$p$-value$\ssymbol{2}$ &$p$-value$\ssymbol{3}$ &$p$-value$\ssymbol{2}$ &$p$-value$\ssymbol{3}$  \\
			\midrule
			VAD & \multirow{5}{*}{DNS} & $0.00^+$&$0.00^+$&$0.00^+$&0.08&$0.00^+$&$0.02^+$\\
			HL-Linear &  & $0.00^+$&$0.00^+$&$0.00^+$&0.06&$0.00^+$&0.05\\
			AMFT-Sequence \& HL-Encoder & & $0.00^+$&$0.00^+$&$0.00^+$&$0.00^+$&$0.00^+$&$0.00^+$\\
			AMFT-Sequence \& HL-Encoder \& VAD & & $0.00^+$&$0.00^+$&$0.00^+$&$0.00^+$&$0.00^+$&$0.00^+$\\
			\midrule
			VAD & \multirow{5}{*}{Libri} & $0.00^+$&$0.03^+$&$0.00^+$&$0.00^+$&$0.00^+$&$0.00^+$\\
			HL-Linear &  & $0.00^+$&$0.00^+$&$0.00^+$&0.09&$0.00^+$&$0.01^+$\\
			AMFT-Sequence \& HL-Encoder & & $0.00^+$&$0.00^+$&$0.00^+$&$0.00^+$&$0.00^+$&$0.00^+$\\
			AMFT-Sequence \& HL-Encoder \& VAD & & $0.00^+$&$0.00^+$&$0.00^+$&$0.00^+$&$0.00^+$&$0.00^+$\\
			\midrule
			\multirow{2.5}{*}{Modules} & 
			\multirow{2.5}{*}{Test set} &
			\multicolumn{2}{c}{SDR} & 
			\multicolumn{2}{c}{SI-SNR} &
			\multicolumn{2}{c}{STOI} \\
			\cmidrule(lr){3-4} \cmidrule(lr){5-6} \cmidrule(l){7-8}
			&& $p$-value$\ssymbol{2}$ &$p$-value$\ssymbol{3}$ &$p$-value$\ssymbol{2}$ &$p$-value$\ssymbol{3}$ &$p$-value$\ssymbol{2}$ &$p$-value$\ssymbol{3}$  \\
			\midrule
			VAD & \multirow{5}{*}{DNS} & $0.00^+$&$0.00^+$&$0.00^+$&$0.00^+$&$0.00^+$&$0.00^+$\\
			HL-Linear & & $0.00^+$&$0.00^+$&$0.00^+$&$0.00^+$&$0.00^+$&$0.00^+$\\
			AMFT-Sequence \& HL-Encoder & & $0.00^+$&$0.00^+$&$0.00^+$&$0.00^+$&$0.00^+$&$0.00^+$\\
			AMFT-Sequence \& HL-Encoder \& VAD & & $0.00^+$&$0.00^+$&$0.00^+$&$0.00^+$&$0.00^+$&$0.00^+$\\
			\midrule
			VAD & \multirow{5}{*}{Libri} & $0.00^+$&0.19&$0.00^+$&0.05&$0.00^+$&$0.00^+$\\
			HL-Linear &  & $0.00^+$&$0.00^+$&$0.00^+$&$0.00^+$&$0.00^+$&$0.00^+$\\
			AMFT-Sequence \& HL-Encoder & & $0.00^+$&$0.00^+$&$0.00^+$&$0.00^+$&$0.00^+$&$0.00^+$\\
			AMFT-Sequence \& HL-Encoder \& VAD & & $0.00^+$&$0.00^+$&$0.00^+$&$0.00^+$&$0.00^+$&$0.00^+$\\
			\bottomrule
		\end{tabular}%
	}
	\vspace{-10pt}
\end{table}

For our ablation study, we develop various variants trained on the DNS-Challenge dataset to investigate the effectiveness of each proposed component.
These variants are built on two base models: AFN-HearNet-base, revised from the widely used FT-Conformer backbone network of CMGAN~\cite{cao2022cmgan}, and HA-MGAN~\cite{cheng2023speech}, the FT-LSTM backbone network that first jointly performs NR and HLC tasks.
Both models utilize the HL-Embedded method~\cite{cheng2023speech} to extend the input hearing loss audiogram.
For each base model, the +HL-Linear variant replaces the HL-Embedded module with linear interpolation to transform the audiogram features, while the +VAD variant incorporates the VAD auxiliary task during training.
The +AMFT-Conformer\&HL-Encoder variant combines the HL-Encoder and AMFT-Conformer with the base model, whereas the +AMFT-LSTM\&HL-Encoder variant integrates the HL-Encoder and AMFT-LSTM that replaces the Conformer module in AMFT-Conformer with an LSTM sequence module. Both variants replace the in-context fusion strategy with our proposed affine modulation fusion strategy.

To ensure the integrity of the experiments, the ablation study is conducted on two test sets aimed at validating the effectiveness of each block and assessing their generalization capabilities across different noisy distributions. 
The first experiment was conducted on the DNS-Challenge validation and test sets, while the second was performed on the LibriSpeech+NoiseX-92 test set.
The detailed results are shown in Table\,\ref{table_ablation_study_dns} and Table\,\ref{table_ablation_study_libri}.
Additionally, Table \ref{table_significance} illustrates the Student's t-test significance analysis results of each proposed component.

\subsubsection{Impact of audiogram linear fitting interpolation for feature alignment}

The audiogram linear interpolation method first extends the sparse audiogram features and aligns them with the spectrum features in terms of frequency components.
By comparing the results of the two baselines to those of the +HL-Linear variants in Table \ref{table_ablation_study_dns} and Table \ref{table_ablation_study_libri}, we can conclude that the linear interpolation method outperforms the HL-Embedded interpolation method across both baselines.
It significantly improves the performance ($p<0.05$) of the FT-LSTM baseline across all metrics on both test sets.
Specifically, there are relative improvements of 10.1\% in HASQI, 7.1\% in WB-PESQ, 4.7\% in NB-PESQ, 17.1\% in SDR, 20.7\% in SI-SNR, and 1.7\% in STOI on the DNS-Challenge test set.
In the case of the FT-Conformer baseline, the linear interpolation method shows marginal significance ($p<0.07$) in improving the WB- and NB-PESQ metrics, along with significant improvements in other metrics.

These findings underscore the importance of the proposed linear interpolation strategy. 
By transforming sparse clinical measurements into dense, spectrum-aligned representations, the HL-Linear method provides a more physiologically plausible approach to audiogram-based feature expansion.
This not only enhances speech quality and intelligibility metrics but also demonstrates the general utility of explicit audiogram interpolation across different neural architectures.

\subsubsection{Impact of affine modulation-based adaptive audiogram fusion}

The results on the DNS-Challenge test set in Table \ref{table_ablation_study_dns} illustrate that both +AMFT-LSTM and +AMFT-Conformer variants significantly outperform their baselines using the in-context fusion strategy ($p<0.05$) across all metrics.
Specifically, the +AMFT-LSTM variant demonstrates notable improvements with HASQI, WB-PESQ, NB-PESQ, SDR, SI-SNR, and STOI metrics increasing by 11.6\%, 10.0\%, 6.7\%, 21.1\%, 25.3\%, and 2.0\%, respectively.
While maintaining the high score of 0.81 in the HASQI metric, the +AMFT-Conformer variant increases the WB-PESQ metric from 2.70 to 2.75, NB-PESQ from 3.23 to 3.27, SDR from 13.57 to 13.92, SI-SNR from 13.33 to 13.68, and STOI from 95.17 to 95.55 compared to its baseline.
Moreover, Table \ref{table_ablation_study_libri} demonstrates the consistency results that both the +AMFT-Conformer and +AMFT-LSTM variants significantly outperform their respective baselines across all SNR conditions and evaluation metrics.

These cross-dataset consistency results confirm the superiority of the affine modulation-based fusion strategy compared to the in-context fusion strategy, demonstrating its effectiveness in enhancing both perceptual quality and intelligibility while maintaining robustness across diverse acoustic conditions.

\begin{figure*}[t]
	\centering
	\includegraphics[width=0.9\linewidth]{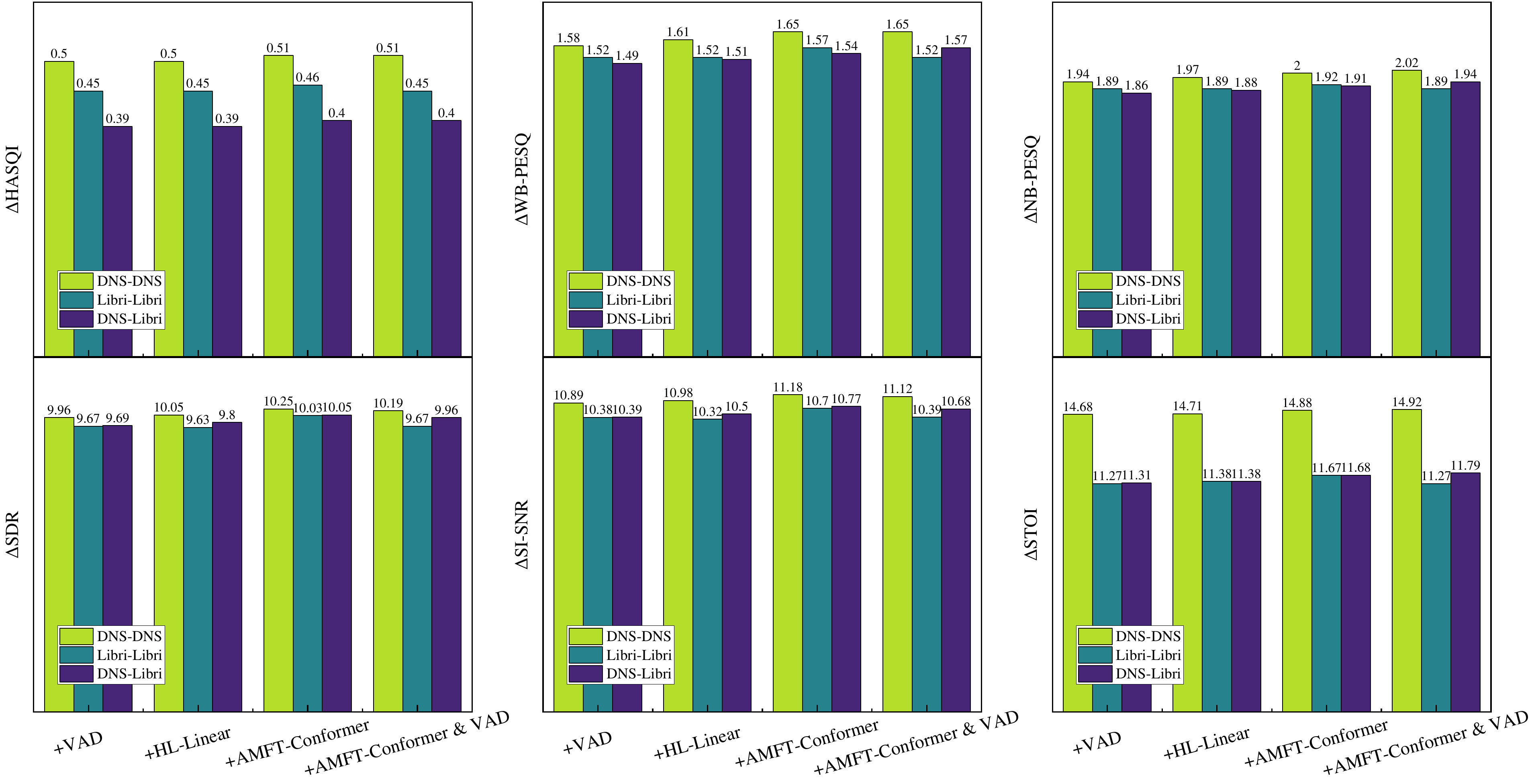}
	\vspace{0.5pt}
	\caption{Results of the ablation study variants under matched and mismatched acoustic conditions. DNS–DNS and Libri–Libri correspond to matched settings, where training and testing are conducted on the DNS-Challenge dataset and the LibriSpeech+DEMAND dataset, respectively. DNS–Libri denotes the double-mismatch condition $\mathcal{C}_{d}$, in which models are trained on DNS-Challenge but evaluated on LibriSpeech+DEMAND dataset.}
	\label{fig_gene}
	\vspace{-0.2cm}
\end{figure*}

\subsubsection{Impact of VAD auxiliary training task}

The VAD auxiliary training task implicitly drives the AMFT-Sequence modules to incorporate speech activity information into the unified deep representation.
It is interesting to find out that the effectiveness of the VAD auxiliary training task varies across the two baselines.
For the FT-LSTM baseline using an in-context fusion strategy, the results of the +VAD variant in Table \ref{table_ablation_study_dns} and Table \ref{table_ablation_study_libri} demonstrate that incorporating the speech activity information significantly increases the performance across all metrics. 
The +VAD variant improves the HASQI, WB-PESQ, NB-PESQ, SDR, SNR, and STOI scores on both the validation and test sets, with notable increases in SDR from 10.11 to 12.18 on the validation set and SI-SNR from 9.08 to 11.48. 
However, the improvements on the FT-Conformer baseline are modest, with SDR increasing only from 13.57 to 13.63 and STOI from 95.17 to 95.31 on the test set. 

The VAD auxiliary training achieves similar performance results with the affine modulation-based feature fusion strategy.
The improvements for the +ACFT-Conforme\&HL-Encoder\&VAD are more modest compared to the +ACFT-Conforme\&HL-Encoder variant.
Table \ref{table_ablation_study_dns} illustrates that on the validation set, NB-PESQ increases from 3.20 to 3.22, and STOI rises from 92.51 to 92.57.
On the test set, NB-PESQ improves from 3.27 to 3.29, and STOI increases from 95.51 to 95.55.
These observations suggest that the Conformer’s self-attention mechanism may already be implicitly learning to differentiate speech from non-speech segments by leveraging its capacity to model long-range dependencies and global context.
Specifically, inter-frame attention computes pairwise similarities between frames via query-key dot products, effectively capturing relationships rooted in diverse acoustic representations and temporal dynamics. 
As a result, after applying a sigmoid activation (or softmax normalization), the attention weights corresponding to speech–speech interactions are considerably higher than those between speech and noise, or between noise frames~\cite{cheng2024residual}.
This inherent characteristic of self-attention may already compensate for the benefits provided by explicit VAD information.
Consequently, incorporating VAD supervision yields relatively smaller performance gains, as its information becomes partially redundant. 
This highlights the strong representational capacity of the Conformer, which reduces the marginal utility of auxiliary cues that overlap with the patterns already captured by the architecture itself.
In summary, the VAD auxiliary training task is a valuable addition, particularly for the simpler FT-LSTM base model, while its utility for the FT-Conformer base model is more limited but still positive.

To further improve the effectiveness of auxiliary supervision, future work may explore tasks that offer more orthogonal information to the Conformer’s internal representations.
For instance, noise type classification or SNR estimation has been investigated in prior works as complementary information for guiding speech enhancement~\cite{may2018signal, kowalewski2020perceptual}. 
These tasks introduce domain-relevant cues that are less likely to be implicitly modeled by the self-attention mechanism, and thus may offer additional improvements when combined with architectures like the Conformer.

\subsubsection{Generalization gap analysis}

To assess the generation gap~\cite{gonzalez2023assessing} of each variant under unseen acoustic conditions, all variants are further trained on the LibriSpeech+DEMAND dataset. 
We define the DNS-Challenge dataset as the double-mismatch acoustic condition, $\mathcal{C}_{d} = \left(\mathcal{S}^{\text{speech}}_{\text{DNS}}, \mathcal{S}^{\text{noise}}_{\text{DNS}}\right)$, where both the speech and noise sources differ from those in the LibriSpeech+Demand test set, i.e., $\mathcal{S}^{\text{speech}}_{\text{DNS}} \neq \mathcal{S}^{\text{speech}}_{\text{LibriSpeech}},\mathcal{S}^{\text{noise}}_{\text{DNS}} \neq \mathcal{S}^{\text{noise}}_{\text{Demand}}$.
The variants trained on LibriSpeech+DEMAND can thus be regarded as reference variants, enabling a clearer separation between the effect of exposure to new data and the effect of increased task difficulty in speech enhancement.

\begin{figure*}[t]
	\setlength{\abovecaptionskip}{-0.2cm}
	\centering
	\includegraphics[width=\linewidth]{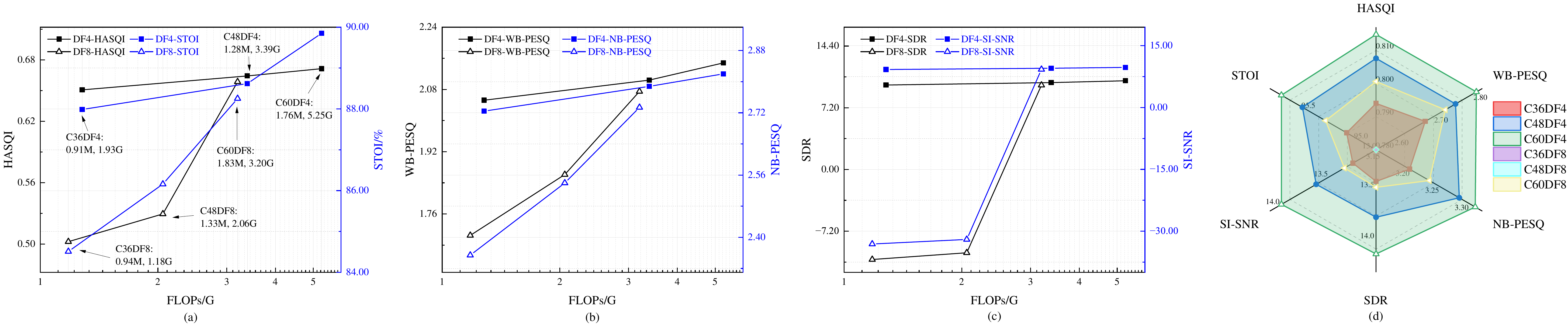}
	\vspace{0.5pt}
	\caption{The impact of different configurations on model performance and complexity, with respect to the number of middle channels and the frequency dimension down-sampling factor. Each line in panels\,(a)-(c) comprises three points representing C36, C48, and C60, respectively, displaying the metrics on the LibriSpeech+NoiseX-92 test set. Meanwhile, panel\,(d) presents the results on the DNS-Challenge test set.}
	\label{fig_complexity}
	\vspace{-0.2cm}
\end{figure*}

\begin{figure}[t]
	\setlength{\abovecaptionskip}{-0.2cm}
	\centering
	\includegraphics[width=0.6\linewidth]{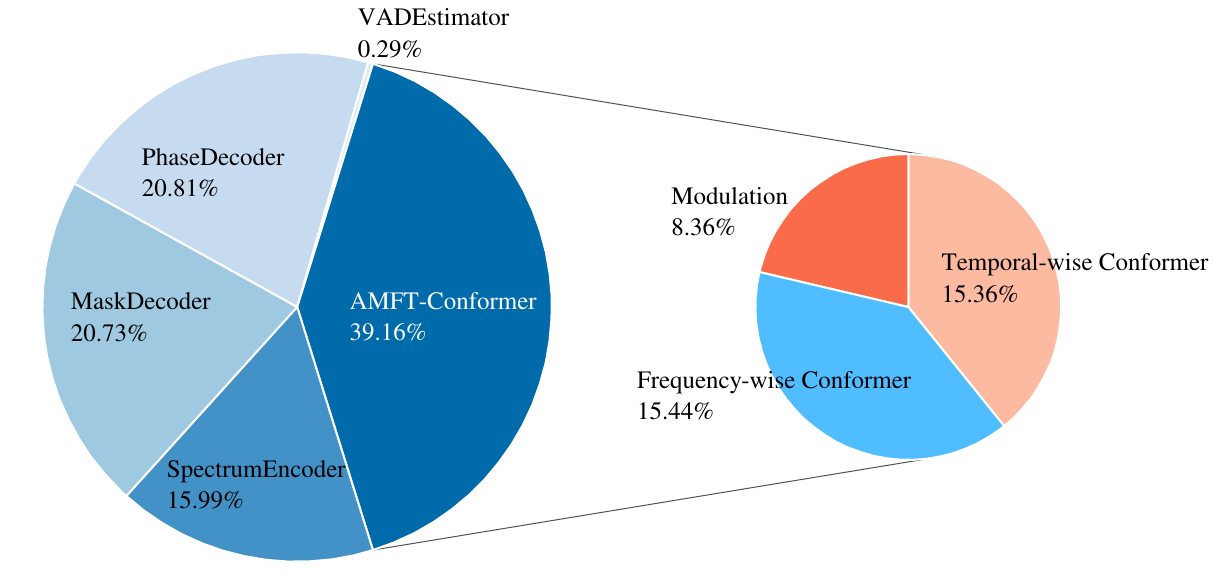}
	\caption{Proportional contribution of MACs from different network modules to the total computational cost.}
	\label{fig_complexity_macs}
	\vspace{-0.2cm}
\end{figure}

The improvements between the unprocessed noisy signal and the enhanced output are denoted as $\Delta$\textit{metric}, such as $\Delta$HASQI and $\Delta$SDR, as illustrated in Fig.\,\ref{fig_gene}.
Across all variants, the matched conditions, i.e., DNS-DNS and Libri-Libri, consistently achieve higher metrics than the mismatched setting  $\mathcal{C}_d$, reflecting the performance degradation caused by condition mismatch. 
Although all variants experience performance drops under speech and noise mismatches, the +AMFT-Conformer \& VAD, i.e., AFN-HearNet, variant maintains relatively higher scores, indicating stronger generalization to unseen acoustic conditions. 
Additionally, the +AMFT-Conformer \& VAD variant is the only one that outperforms its corresponding reference model across all metrics on the mismatched test set, with the exception of HASQI. 
This is evidenced by the generation gap $G_E = \tfrac{\Delta E - \Delta E_{\text{ref}}}{\Delta E_{\text{ref}}}$, yielding –11.11\% for HASQI, 3.29\% for WB-PESQ, 2.65\% for NB-PESQ, 3.00\% for SDR, 2.79\% for SI-SNR, and 4.61\% for STOI.

These findings suggest that although the VAD auxiliary training task does not yield substantial improvements for the Conformer backbone under matched conditions, it nevertheless provides complementary benefits by enhancing robustness under mismatched conditions.
This advantage may be attributed to the fact that the Conformer primarily learns speech distribution patterns through implicit learning during sequence modeling, whereas the VAD module leverages explicit supervision to discriminate speech from noise, thereby yielding stronger generalization capability across unseen domains.

\subsubsection{Complexity analysis}

For the HAs system, there is a crucial trade-off between performance and model statistical characteristics, including model size and complexity. 
The computational complexity and parameter size of the AFN-HearNet are primarily influenced by two key factors: the number of intermediate channels and the dimensionality of the downsampled frequency bins. 
The intermediate channels determine the feature processing dimension within the AMFT-Conformer block, with higher dimensions necessitating greater computational and memory resources. 
Additionally, the dimensionality of the downsampled frequency bins directly affects both the temporal steps in the F-Conformer and the number of parallel processing units in the T-Conformer.

\begin{table}[t]
	\renewcommand{\arraystretch}{1}
	\caption{Results of various baselines on the DNS-Challenge test set.}
	\label{table_result_dns}
	\centering
	\resizebox{\linewidth}{!}{
		\begin{tabular}{l*{8}{>{\centering\arraybackslash}p{0.8cm}}}
			\toprule
			Method & \#Param (M) & FLOPs (G/s) &  HASQI & WB-PESQ & NB-PESQ & SDR (dB) &SI-SNR & STOI (\%)  \\
			\midrule
			Unprocessed & - & - &0.61&1.09&1.29&3.74&2.87&84.77 \\ %
			\midrule
			WDRC-FIG6~\cite{killion19933} &-&-&0.62&1.49&1.98&8.27&8.15&91.52\\
			CRN~\cite{tan2018convolutional} &13.06&1.04&0.72&2.02&2.63&12.44&12.11&91.90\\
			DCCRN~\cite{hu2020dccrn} &3.67&5.63 &0.74&2.23&2.88&13.25&12.83&93.40 \\
			CMGAN~\cite{cao2022cmgan} &0.67&6.59&0.74&1.88&2.42&13.60&13.33&93.58 \\
			NUNet\_TLS~\cite{hwang2022monoaural} & 2.82 & 5.48 &0.78&2.13&2.79&\bf 13.98&\bf 13.71&93.98\\
			CompNet~\cite{fan2023compnet} &3.87&7.13&0.74&1.89&2.45&13.08&12.65&93.53 \\
			HA-MGAN~\cite{cheng2023speech} &0.71 & 2.64 & 0.77&2.66&3.21&11.67&11.11&93.90 \\
			\midrule
			AFN-HearNet-small (proposed) &0.91&1.93&0.79&2.68&3.21&13.51&13.24&95.13   \\ 
			AFN-HearNet (proposed) &1.28&3.39&\bf 0.81&\bf 2.75&\bf 3.29&13.86&13.62& \bf 95.55   \\ 
			\bottomrule
		\end{tabular}
	}
	\vspace{-0.2cm}
\end{table}

\begin{table}[t]
	\renewcommand{\arraystretch}{1}
	\caption{Results of the statistical analysis based on two-tailed Student's t-test on the DNS-Challenge test set for each baseline with respect to AFN-HearNet.} 
	\label{table_result_significance_dns}
	\centering
		\begin{tabular}{lccccccc}
			\toprule
			Method &  HASQI & WB-PESQ & NB-PESQ & SDR  &SI-SNR & STOI  \\
			\midrule
			WDRC-FIG6~\cite{killion19933}  &\plevelstar{3}&\plevelstar{3}&\plevelstar{3}&\plevelstar{3}&\plevelstar{3}&\plevelstar{3}\\
			CRN~\cite{tan2018convolutional} &\plevelstar{3}&\plevelstar{3}&\plevelstar{3}&\plevelstar{3}&\plevelstar{3}&\plevelstar{3}\\
			DCCRN~\cite{hu2020dccrn}    &\plevelstar{3}&\plevelstar{3}&\plevelstar{3}&\plevelstar{1} (0.13)&\plevelstar{2} (0.05)&\plevelstar{3}\\
			CMGAN~\cite{cao2022cmgan} &\plevelstar{3}&\plevelstar{3}&\plevelstar{3}&\plevelstar{1} (0.49)&\plevelstar{1} (0.45)&\plevelstar{3}\\
			NUNet\_TLS~\cite{hwang2022monoaural} &\plevelstar{3}&\plevelstar{3}&\plevelstar{3}&\plevel{1} (0.78)&\plevel{1} (0.84)&\plevelstar{3}\\
			CompNet~\cite{fan2023compnet} &\plevelstar{3}&\plevelstar{3}&\plevelstar{3}&\plevelstar{3}&\plevelstar{3}&\plevelstar{3}\\
			HA-MGAN~\cite{cheng2023speech} &\plevelstar{3}&\plevelstar{1} (0.26)&\plevelstar{1} (0.33)&\plevelstar{3}&\plevelstar{3}&\plevelstar{3}\\
			\bottomrule
		\end{tabular}	
		\begin{tablenotes}
			\item[] The symbols \plevel{1}/\plevelstar{1}, \plevel{2}/\plevelstar{2}, and \plevel{3}/\plevelstar{3} refer to significance levels of $(p>0.1)$, $(0.05<p<0.1)$, and $(p<0.05)$, respectively. Specifically, \plevelstar{1} indicates that AFN-HearNet outperforms the baseline, while \plevel{1} indicates the opposite.
		\end{tablenotes}
		\vspace{-0.2cm}
	\end{table}

To evaluate the model's sensitivity to these factors and gain insights into the trade-offs between model complexity and performance, we conducted experiments with different configurations, denoted as C$m$DF$n$, derived from the Cartesian product of middle channels $m\in[32, 48, 60]$ and frequency dimension down-sampled factor $n\in[4, 8]$.
For example, C36DF8 refers to the model with 36 intermediate channels and frequency down-sampled by a factor of 8, achieved by doubling the number of stacked down-sampling layers in the spectrum encoder.
The results in Fig.\,\ref{fig_complexity}(a)-(c) demonstrate that as the number of channels increases from C36 to C60 in the DF4 configuration, the parameters rise from 0.91\,M to 1.76\,M, representing a 1.9$\times$ increase, while the floating point operations per second (FLOPs) increase from 1.93\,G to 5.25\,G, reflecting a 2.7$\times$ surge.
Similarly, in the DF8 configuration, this channel expansion rises parameters from 0.94\,M to 1.83\,M, also a 1.9$\times$ increment, and FLOPs from 1.18\,G to 3.20\,G, marking a 2.7$\times$ increment.
Additionally, performance metrics indicate that this increase in complexity yields diminishing returns.
Fig.\,\ref{fig_complexity}(d) illustrates that while C60DF4 achieves the highest scores across all metrics, C48DF4 manages to retain 90\% of these scores with 40\% fewer FLOPs.
Although the DF8 variants consistently demonstrate better efficiency, requiring 39\% fewer FLOPs than the DF4 variants with the same number of channels, there are substantial decreases in all metrics, which is unacceptable.
Therefore, adjusting the channel dimension is a more effective way to reduce model complexity.
Consequently, we choose the C48DF4 configuration as the AFN-HearNet, and the C36DF4 variant as the AFN-HearNet-small, given its performance at a lower complexity level.

To further analyze the computational characteristics of the proposed model, we decompose the multiply–accumulate operations (MACs) by major network modules and present their proportional contributions to the overall computational cost, as shown in Fig.\,\ref{fig_complexity_macs}.
Under the C48D4 configuration (i.e., AFN-HearNet), 29.28\% of the total parameters and 39.16\% of the total MACs are contributed by the intermediate AMFT-Conformer module. 
Within this module, the frequency–temporal modeling Conformer accounts for 30.8\% of the overall computational complexity, while the embedded modulation fusion submodule contributes 8.36\%. 
Furthermore, the analysis shows that although the dilated DenseNet exhibits strong global feature extraction capabilities, it is a computation-intensive module, accounting for 15.13\% of the total computational cost despite having relatively few parameters.
Consequently, the dilated DenseNet-based SpectrumEncoder, MaskDecoder, and PhaseDecoder contribute 15.99\%, 20.73\%, and 20.81\% of the total MACs, respectively.
In this work, we propose a personalized speech enhancement model operating under causal constraints with a theoretical latency of 16\,ms, prioritizing performance while maintaining flexibility in optimizing the core computational load. 
Specifically, the modulation fusion can be decoupled from the FT-modeling module, and the Conformer component can be substituted with a more lightweight alternative.
In addition, model compression can be further achieved through knowledge distillation, enabling a lightweight student model to learn from the original network, which may achieve a 2–4× reduction in model size with only a slight degradation in quality~\cite{thakker2022fast}.

\subsection{Comparisons with advanced approaches}

\begin{figure}[t]
	\setlength{\abovecaptionskip}{-0.2cm}
	\centering
	\includegraphics[width=0.9\linewidth]{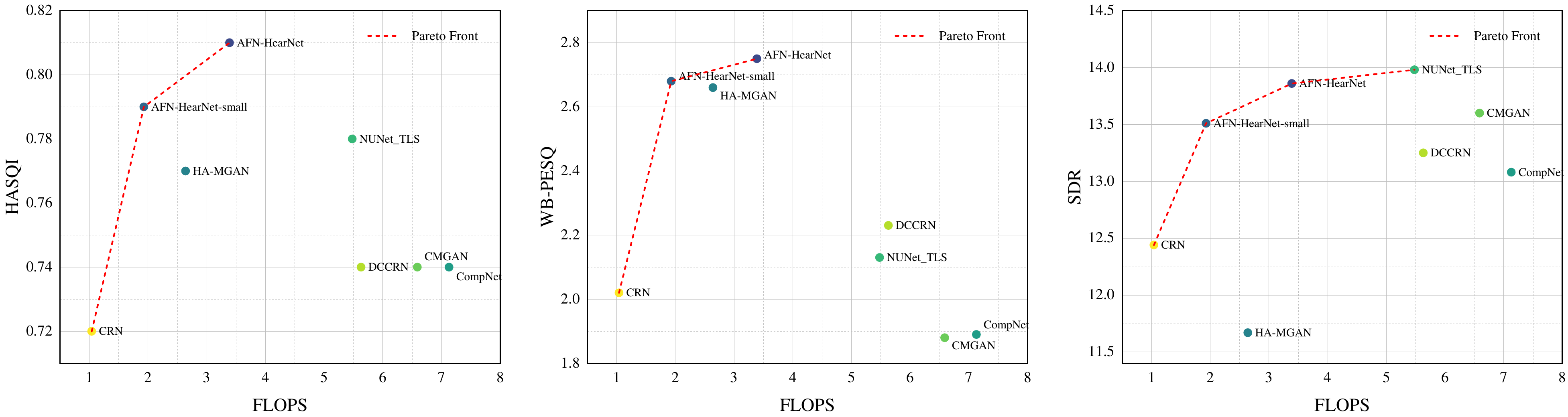}
	\caption{Performance–efficiency trade-offs for different models on the DNS-Challenge test set, showing the relationship between GFLOPs and HASQI, WB-PESQ, and SDR. The red dashed line indicates the Pareto frontier.}
	\label{fig_tradeoff_dns}
	\vspace{-0.2cm}
\end{figure}

\begin{figure*}[t]
	\setlength{\abovecaptionskip}{-0.2cm}
	\centering
	\includegraphics[width=\linewidth]{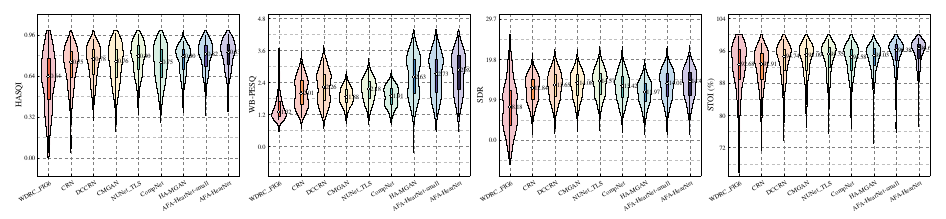}
	\vspace{0.5pt}
	\caption{Metric results from various joint models on the DNS-Challenge test set. The median is indicated by empty diamonds within the violin plots.}
	\label{fig_violin}
	\vspace{-0.2cm}
\end{figure*}

\subsubsection{Results on DNS-challenge dataset}

\begin{table}[t]
\renewcommand{\arraystretch}{1}
\caption{Results of diffusion baseline on the DNS-Challenge test set. $\ssymbol{2}$ denotes the performance of normal speech enhancement.}
\label{table_result_dns_diffusion}
\centering
	\begin{tabular}{l*{8}{>{\centering\arraybackslash}p{0.8cm}}}
			\toprule
			Method & \#Param (M) & FLOPs (G/s) &  HASQI & WB-PESQ & NB-PESQ & SDR (dB) &SI-SNR & STOI (\%)  \\
			\midrule
			Unprocessed & - & - &0.61&1.09&1.29&3.74&2.87&84.77 \\ %
			\midrule
			SGMSE$\ssymbol{2}$~\cite{richter2023speech} & \multirow{2}{*}{65.6} & \multirow{2}{*}{-} & - & 2.77 & 3.31 & 16.73 & 16.61 & 96.83  \\ 
			SGMSE+FIG6~\cite{richter2023speech} &&&\bf 0.83&2.56&3.17&\bf 15.67&\bf 15.48&\bf 96.28 \\
			AFN-HearNet (proposed) &1.28&3.39&0.81&\bf 2.75&\bf 3.29&13.86&13.62& 95.55   \\ 
			\bottomrule
		\end{tabular}
	\vspace{-0.2cm}
\end{table}

Fig.\,\ref{fig_violin} presents the violin plots of results for all baselines on the DNS-Challenge test set, effectively illustrating the distribution of objective metrics across different methods.
The plots indicate that AFN-HearNet not only achieves a better central tendency but also exhibits a relatively narrow distribution compared to all baselines.
Table \ref{table_result_dns} shows the mean results of each model on the DNS-Challenge test set.
The results demonstrate that the AFN-HearNet achieves the highest scores across key speech quality metrics, including HASQI, WB-PESQ, NB-PESQ, and STOI, and considerably outperforms the second-best baseline, HA-MGAN, with improvements of 5.19\%, 3.38\%, 2.49\%, and 1.76\%, respectively. 
While the NUNet\_TLS achieves marginally higher scores in signal-level metrics, with an SDR of 13.98 and an SI-SNR of 13.71, the proposed AFN-HearNet demonstrates competitive performance with an SDR of 13.86 and an SI-SNR of 13.62.
Moreover, the AFN-HearNet delivers substantial improvements across all other key metrics. 
Specifically, the AFN-HearNet outperforms the NUNet\_TLS by 3.85\% in HASQI, achieving 0.81 compared to 0.78; 29.11\% in WB-PESQ, with 2.75 compared to 2.13; 17.92\% in NB-PESQ, with 3.29 compared to 2.79; and 1.67\% in STOI, achieving 95.55 compared to 93.98.
Additionally, the AFN-HearNet exhibits advantages in parameter count and computational complexity, requiring only 45\% and 62\% of the resources used by NUNet\_TLS, respectively, highlighting the superiority of our model compared to NUNet\_TLS.

A two-tailed Student’s t-test is employed to rigorously assess the statistical significance of the observed performance differences between AFN-HearNet and each baseline method, with the corresponding results reported in Table \ref{table_result_significance_dns}.
Overall, AFN-HearNet significantly outperforms all baselines in most metrics, especially in perceptual quality scores, such as HASQI, WB-PESQ, NB-PESQ, and STOI, where significant improvements ($p<0.05$) are observed across nearly all comparisons. 
For the SDR and SI-SNR metrics, AFN-HearNet achieves statistically significant improvements over most methods; however, for a few methods (e.g., DCCRN, CMGAN, and NUNet\_TLS), the differences are not statistically significant ($p > 0.1$), indicating negligible performance gaps.

Fig.\,\ref{fig_tradeoff_dns} illustrates the performance–efficiency trade-off among all compared methods, where the horizontal axis represents computational cost in GFLOPs and the vertical axis denotes the metric scores.
The red dashed line highlights the Pareto frontier, indicating methods that achieve the best possible performance for a given computational budget. 
Both AFN-HearNet and its small variant lie on the Pareto frontier, demonstrating their competitive efficiency–performance profiles.
Although NUNet\_TLS also appears on the Pareto frontier for the SDR metric, it requires 61.65\% more computational resources than AFN-HearNet, yet offers only a negligible SDR gain of 0.12\,dB

\begin{table}[t]
	\caption{Results of various baselines on the LibriSpeech+Demand test set.}
	\label{table_result_libri}
	\centering
		\begin{tabular}{l*{8}{>{\centering\arraybackslash}p{0.8cm}}}
			\toprule
			
			Method & \#Param (M) & FLOPs (G/s) &  HASQI & WB-PESQ & NB-PESQ & SDR (dB) & SI-SNR & STOI (\%)  \\
			\midrule
			Unprocessed & - & - & 0.30 & 1.10 &1.27& 3.67 &2.50& 80.63 \\ %
			\midrule
			DB-AIAT~\cite{yu2022dual} &1.05&13.12&0.71&1.79&2.37&\bf 14.32&\bf 13.86&90.17 \\
			MP-SENet~\cite{lu2023mp} &2.13&19.46&0.69&2.18&2.88&10.89&9.63&91.46\\
			FSPEN~\cite{yang2024fspen} &0.03&0.08&0.61&1.94&2.44&9.75&8.95&86.26\\
			SEMamba~\cite{chao2024investigation} &1.18&12.74&0.68&2.35&2.95&10.95&9.64&91.70\\
			PrimeK-Net~\cite{lin2025primek} &1.41&11.18&0.72&2.62&3.20&11.46&10.12&\bf 93.06 \\
			\midrule
			AFN-HearNet-small (proposed) &0.91&1.93&0.75&2.57&3.12&13.15&12.63&91.65  \\ 
			AFN-HearNet (proposed) &1.28&3.39&\bf 0.77&\bf 2.67&\bf 3.20&13.72&13.28&92.35   \\ 
			\bottomrule
		\end{tabular}
	\vspace{-0.2cm}
\end{table}

\begin{table}[t]
\renewcommand{\arraystretch}{1}
\caption{Results of the statistical analysis based on two-tailed Student's t-test on the LibriSpeech+Demand test set for each baseline with respect to AFN-HearNet.} 
\label{table_result_significance_libri}
\centering
	\begin{tabular}{lccccccc}
		\toprule
		Method &  HASQI & WB-PESQ & NB-PESQ & SDR  &SI-SNR & STOI  \\
		\midrule
		DB-AIAT~\cite{yu2022dual}  &\plevelstar{3}&\plevelstar{3}&\plevelstar{3}&\plevel{3}&\plevel{3}&\plevelstar{3}\\
		MP-SENet~\cite{lu2023mp} &\plevelstar{3}&\plevelstar{3}&\plevelstar{3}&\plevelstar{3}&\plevelstar{3}&\plevelstar{3}\\
		FSPEN~\cite{yang2024fspen}  &\plevelstar{3}&\plevelstar{3}&\plevelstar{3}&\plevelstar{3}&\plevelstar{3}&\plevelstar{3}\\
		SEMamba~\cite{chao2024investigation} &\plevelstar{3}&\plevelstar{3}&\plevelstar{3}&\plevelstar{3}&\plevelstar{3}&\plevelstar{3}\\
		PrimeK-Net~\cite{lin2025primek}&\plevelstar{3}&\plevelstar{1} (0.11)&\plevelstar{1} (0.69)&\plevelstar{3}&\plevelstar{3}&\plevel{3}\\
		\bottomrule
	\end{tabular}	
	\vspace{-0.2cm}
\end{table}

Unlike above masking- or mapping-based speech enhancement baselines that directly estimate clean spectral components from noisy inputs, the diffusion model~\cite{lu2021study,richter2023speech,richter2024causal} adopts a fundamentally different generative paradigm, reconstructing enhanced speech through iterative denoising.
We further train the SGMSE~\cite{richter2023speech} diffusion model on the DNS-Challenge dataset as a standard speech enhancement module, followed by the FIG6 HLC algorithm to achieve personalized speech enhancement. 
This two-stage configuration conforms to the classic hearing-aid compensation framework, enabling evaluation of a diffusion model–based approach within this established paradigm. 
Under conditions of a large model footprint and non-causal processing, diffusion models tend to deliver superior performance, benefiting from their iterative refinement process and strong generative prior. 
However, by comparing the results of SGMSE and SGMSE+FIG6, we observe that although SGMSE alone achieves strong denoising performance, the subsequent application of the HLC algorithm consistently degrades all quality metrics. 
Notably, WB/NB PESQ scores decrease from 2.77/3.31 to 2.56/3.17. 
This degradation suggests that the two-stage compensation paradigm, in which speech enhancement and hearing compensation operate as decoupled processes, fails to preserve the intricate interdependencies between noise suppression and perceptual compensation, thereby introducing additional distortions. 
In contrast, the jointly optimized AFN-HearNet demonstrates a clear advantage in speech quality, indicating that integrating enhancement and compensation within a unified framework better preserves perceptual fidelity.

\subsubsection{Results on LibriSpeech+Demand dataset}

\begin{figure}[t]
	\setlength{\abovecaptionskip}{-0.2cm}
	\centering
	\includegraphics[width=0.9\linewidth]{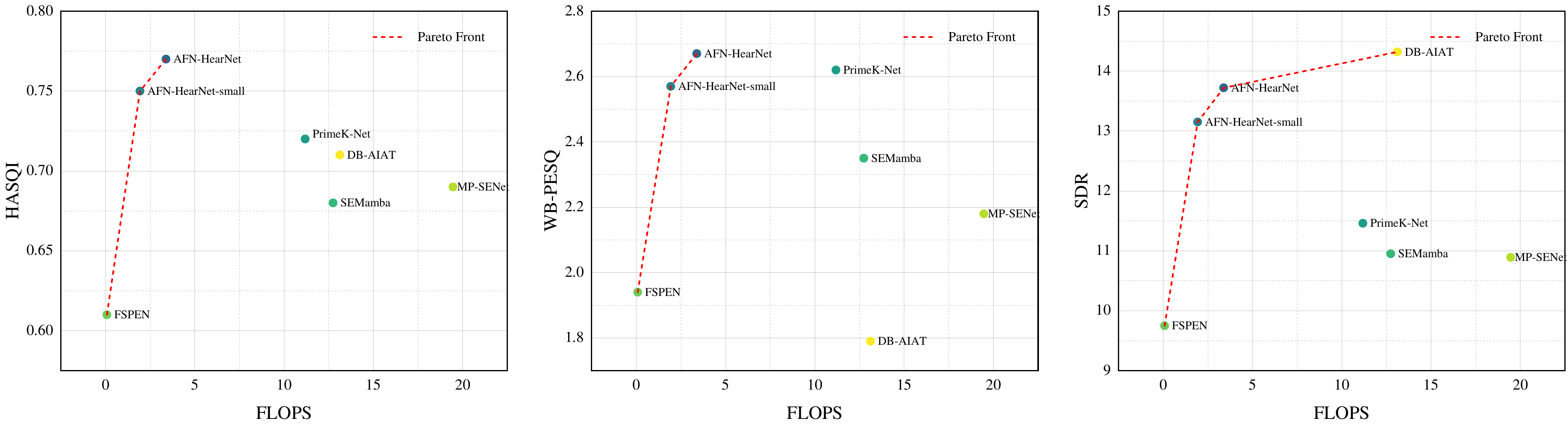}
	\caption{Performance–efficiency trade-offs for different baselines on the LibriSpeech+Demand test set.}
	\label{fig_tradeoff_libri}
	\vspace{-0.2cm}
\end{figure}

\begin{figure*}[t]
	\setlength{\abovecaptionskip}{-0.2cm}
	\centering
	\includegraphics[width=\linewidth]{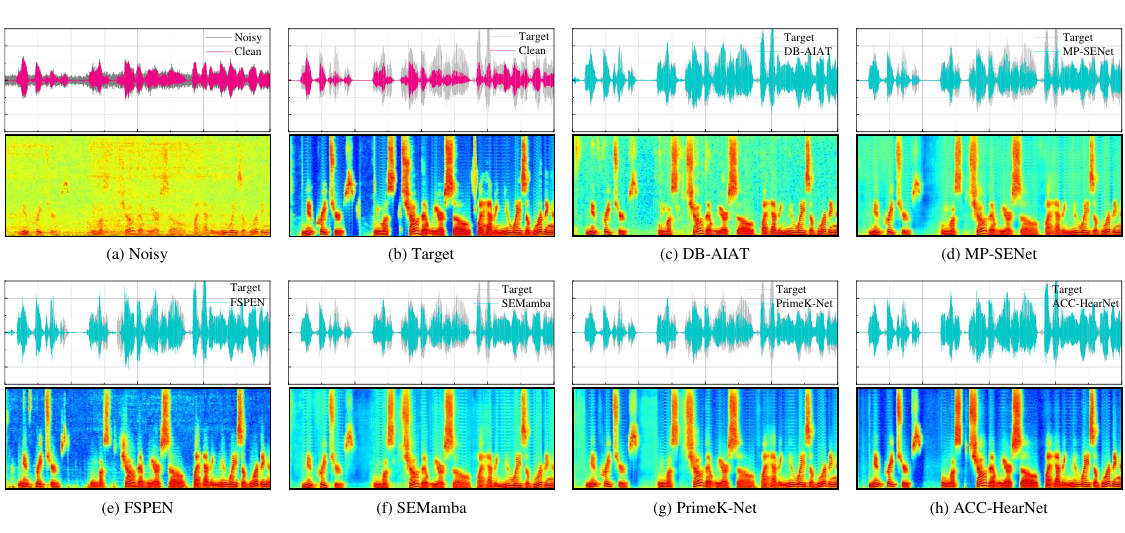}
	\caption{Comparison of waveform and spectrum enhancement results from various joint models on the LibriSpeech+Demand test set.}
	\label{fig_results_on_libri}
	\vspace{-0.2cm}
\end{figure*}

Fig.\,\ref{fig_results_on_libri} visualizes the results of various models on the LibriSpeech+Demand test set, demonstrating that our AFN-HearNet and DB-AIAT joint model outperform others in terms of speech compensation.
Additionally, the AFN-HearNet shows better performance than DB-AIAT in the NR task.
This finding aligns with Table \ref{table_result_libri}, which illustrates the detailed results of various baselines on the LibriSpeech+Demand test set.
The statistical analysis confirms that AFN-HearNet achieves significant improvements ($p$ < 0.05) over all baselines in HASQI, WB-PESQ, and NB-PESQ, with the exception of PrimeK-Net, for which the differences are not statistically significant ($p$ > 0.1).
Although the DB-AIAT baseline achieves the highest scores in SDR at 14.32 and SI-SNR at 13.86, the AFN-HearNet closely follows with scores of 13.72 and 13.28.
Despite the marginal difference in SDR and SI-SNR, the AFN-HearNet demonstrates considerable overall performance across other metrics, highlighting its competitiveness.
Specifically, in terms of HASQI, WB-PESQ, and NB-PESQ, the AFN-HearNet significantly outperforms the DB-AIAT by 8.45\%, 49.16\%, and 35.02\%, respectively.
Compared to the PrimeK-Net's STOI performance of 93.06, the AFN-HearNet maintains a comparable intelligibility score of 92.35 and excels in other key metrics, achieving improvements of 6.94\% in HASQI, 1.91\% in WB-PESQ, 19.72\% in SDR, and 31.23\% in SI-SNR.

These results demonstrate that the AFN-HearNet offers a more comprehensive performance with lower computational costs.
Specifically, it attains the highest HASQI and WB-PESQ scores while requiring only 3.39 G FLOPs, which is less than one-third of PrimeK-Net (11.18\,G) and approximately one-quarter of DB-AIAT (13.12\,G).
This corresponds to computational cost reductions of 69.69\% and 74.16\%, respectively
Additionally, Fig.\,\ref{fig_tradeoff_libri} illustrates the performance–efficiency trade-offs of AFN-HearNet and the baselines on the LibriSpeech test set, with FLOPs plotted against HASQI, WB-PESQ, and SDR. 
Both AFN-HearNet and its small variant occupy positions along the Pareto frontier.
These results demonstrate that the AFN-HearNet offers a more comprehensive performance with lower computational costs.

\section{Conclusion}
\label{section_conclusion}

In this paper, we propose a novel joint model named AFN-HearNet that delivers PSE tailored to the individual's hearing capabilities.
By fusing cross-domain features of the spectrum and the hearing loss audiogram, the AFN-HearNet tackles the NR and HLC tasks simultaneously, effectively addressing the interactions between these two tasks and enabling systematic optimization.
We propose an audiogram-specific encoder that transforms the sparse audiogram features into personalized deep representations, addressing the alignment problem in cross-domain feature fusion.
Following the encoders, the AMFT-Conformer integrates affine modulation and Conformer's attention mechanisms, iteratively fusing the spectrum deep representations and the audiogram-derived personalized representations at different levels.
In the speech reconstruction stage, we introduce a VAD auxiliary training task that implicitly embeds speech and non-speech patterns into the unified deep representations, which helps the model better distinguish between speech and non-speech segments, further improving the overall performance.
Experimental results show that the AFN-HearNet significantly outperforms SOTA in-context fusion joint approaches across the key metrics of HASQI and PESQ.
Moreover, it maintains a low parameter count and computational complexity, representing a considerable trade-off between performance and efficiency.
In future work, we plan to explore extensions for binaural HA scenarios by leveraging spatial audio cues to enhance the listening experience. 
Additionally, we aim to investigate knowledge distillation techniques to reduce model complexity, making it more suitable for resource-constrained devices. 

\section{Acknowledgments}
\label{ack}

This work was supported in part by the National Key Research and Development Program of China under Grant No.\,2020YFC2004003, the National Natural Science Foundation of China under Grant No.\,61871213, and by the China Scholarship Council under Grant No.\,202406090186.

\appendix

\section*{Extra information}
\section{Impact of different weighting factors on generative loss}
\label{appdix_a}

Among the generative loss function defined in Equation~\eqref{loss_eqa}, the adversarial loss $\mathcal{L}_{adv}$ and the perceptual loss $\mathcal{L}_{pmsqe}$ aim to jointly optimize the key metrics HASQI and PESQ, whose performance is directly influenced by the weighting factors (i.e., $\alpha$ and $\lambda$).
We evaluate different weight combinations by first fixing $\alpha=1.00$ with $\lambda \in \{0.30, 0.45, 2.00\}$, and then fixing $\lambda=0.30$ with $\alpha \in \{0.50, 1.00, 1.50, 2.50\}$.
Results in Table \ref{table_ablation_weights} indicate that, when fixing the HASQI adversarial loss weight $\alpha=1.00$, increasing the PMSQE loss weight $\lambda$ leads to consistent improvements in WB/NB-PESQ, SDR, and SI-SNR metrics, but with a slight decrease in the STOI metric.
These results demonstrate that emphasizing the PMSQE perceptual loss facilitates improvements in perceptual quality. Nevertheless, they also highlight that the optimization objectives for perceptual speech quality and intelligibility are not inherently aligned.
As $\alpha$ increases from 0.5 to 1.5, the evaluation metrics remain relatively stable. 
Nevertheless, a substantial degradation is observed at $\alpha=2.5$, suggesting that an overly large adversarial loss weight adversely affects the overall reconstruction performance.

\begin{table}[t]
	\renewcommand{\arraystretch}{1}
	\caption{Performance of models trained with GAN loss on the DNS-Challenge dataset using different loss weight configurations.}
	\label{table_ablation_weights}
	\centering
		\begin{tabular}{llcccccc}
			\toprule
			\multicolumn{2}{c}{Weight Factors} &  HASQI & WB-PESQ & NB-PESQ  &SDR & SI-SNR & STOI  \\
			\midrule
			  &$\lambda=0.30$   & 0.81&2.75&3.27 & 13.92 & 13.68 & 95.51 \\
			$\alpha=1.00$&$\lambda=0.45$ &0.81&2.78&3.29 & 13.95 & 13.71 & 95.49 \\
					  &$\lambda=2.00$   & 0.81&2.84&3.35 & 13.64 & 13.33 & 95.42 \\
			\midrule
			$\alpha=0.50$ & \multirow{4}{*}{$\lambda=0.30$}&0.81&2.76&3.28 & 13.92 & 13.66 & 95.49\\
			$\alpha=1.00$  && 0.81&2.75&3.27 & 13.92 & 13.68 & 95.51 \\
			$\alpha=1.50$ && 0.81 &2.76&3.28 & 13.98 & 13.72 & 95.52 \\
			$\alpha=2.50$ && 0.80  & 2.70&3.24 & 13.79 & 13.51 & 95.31 \\
			\bottomrule
		\end{tabular}
\end{table}


\bibliographystyle{cas-model2-names}

\bibliography{main}

\clearpage

\bio{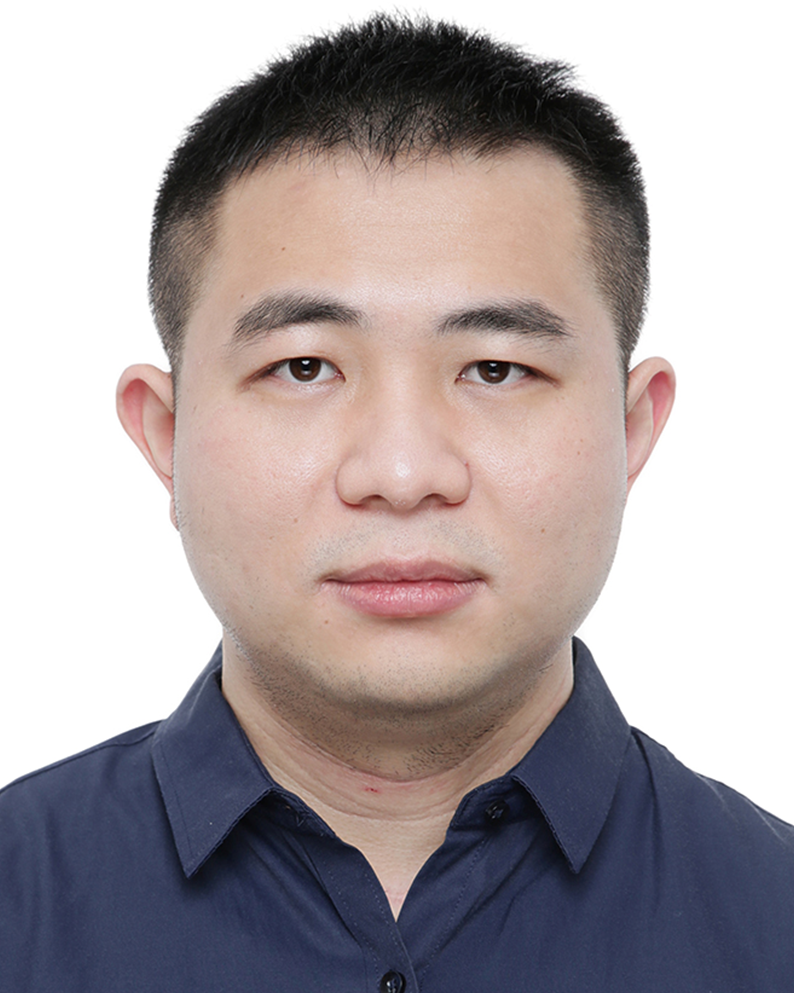}
\textbf{Ye Ni} received the M.S.\ degree from Nanjing University, Nanjing, China, in 2022. He is currently working toward a Ph.D.\ degree from Southeast University, Nanjing, China. His research interests include deep learning-based speech enhancement, acoustic echo cancellation, and signal processing.
\endbio
\vskip6pc

\bio{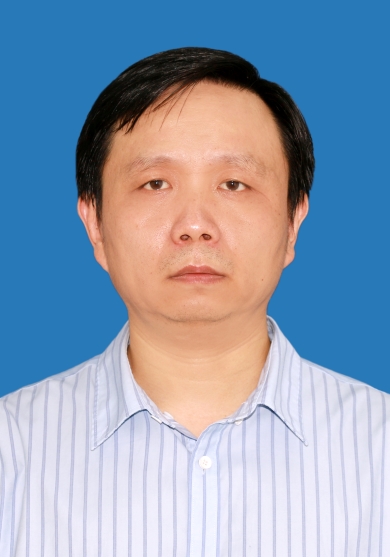}
\textbf{Ruiyu Liang} (Member, IEEE) received the Ph.D.\ degree from Southeast University, Nanjing, China, in 2012. He is a Professor at the Nanjing Institute of Technology, Nanjing, China. His research interests include speech signal processing and signal processing for hearing aids.
\endbio
\vskip6pc

\bio{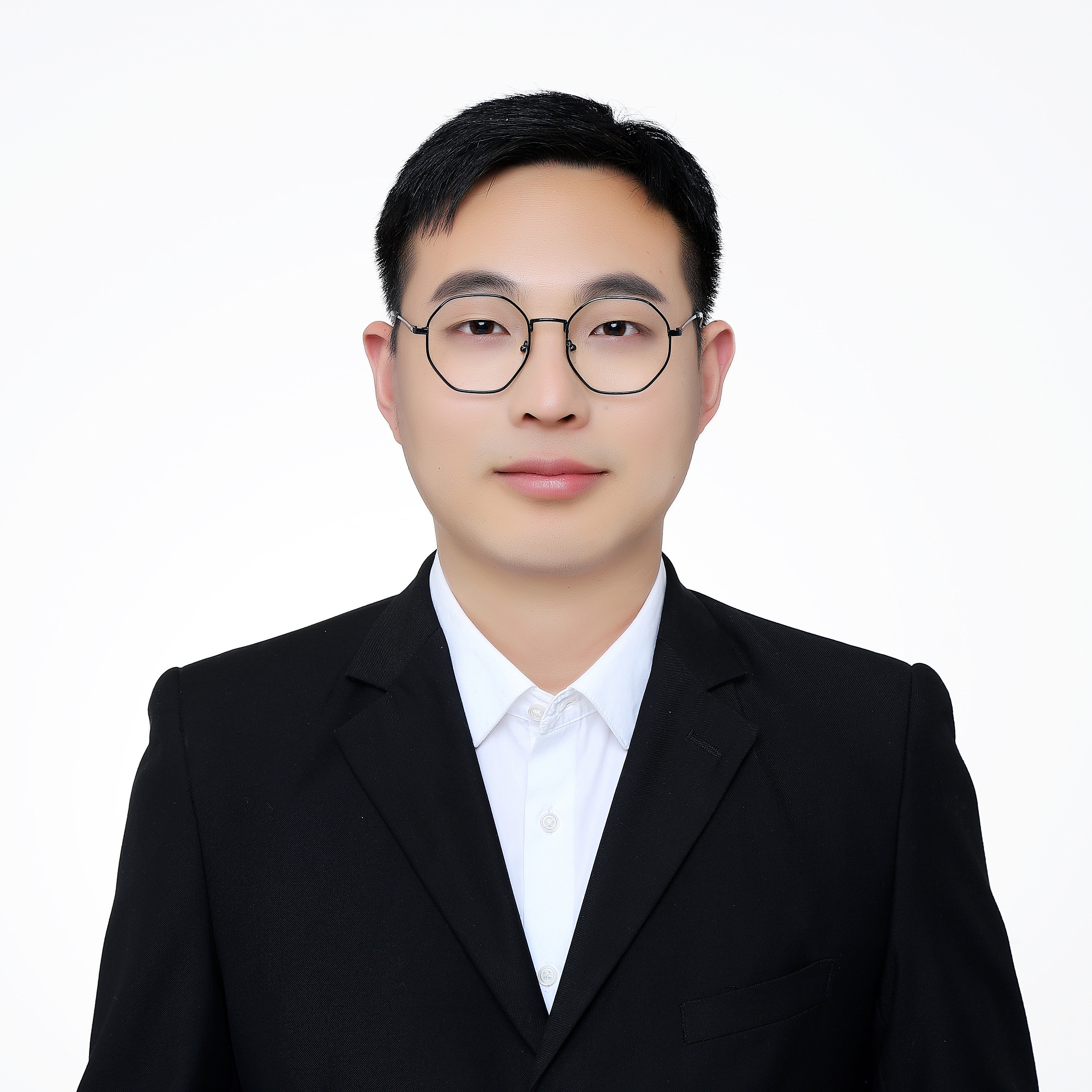}
\textbf{Xiaoshuai Hao} received his Ph.D. from the Institute of Information Engineering, Chinese Academy of Sciences, in 2023. He is currently a research expert in multimodal algorithms for autonomous driving and robotics at Xiaomi EV. His research interests include embodied intelligence, multimodal learning, and autonomous driving. Dr. Hao has published over 30 papers in top-tier journals and conferences, including TIP, Information Fusion, NeurIPS, ICLR, ICML, CVPR, ICCV, ECCV, ACL, AAAI, and ICRA. He has achieved significant success in international competitions, securing top-three placements at prestigious conferences such as CVPR and ICCV. Additionally, he serves on the editorial board of Data Intelligence and is an organizer for the RoDGE Workshop at ICCV 2025 and the RoboSense Challenge at IROS 2025.
\endbio
\vskip6pc

\bio{jiaming.jpg}
\textbf{Jiaming Cheng} received the B.E.\ degree from Nanjing Normal University, Nanjing, China, in 2019. He is currently working toward a Ph.D.\ degree from Southeast University, Nanjing, China. His research interests include speech enhancement and deep learning.
\endbio
\vskip6pc

\bio{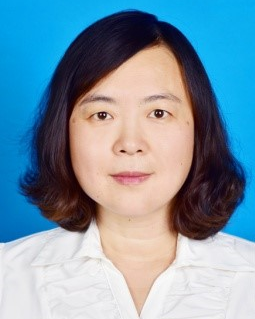}
\textbf{Qingyun Wang} Qingyun Wang received the M.S. degree in Computer Engineering and the Ph. D. degree in Information and Communication Engineering from Southeast University, Nanjing, China, in 2001 and 2011 respectively. Now she is a professor of Nanjing Institute of Technology, Nanjing, China. Her current research interests include acoustic signal processing, speech enhancement and microphone array signal processing.
\endbio
\vskip6pc

\clearpage

\bio{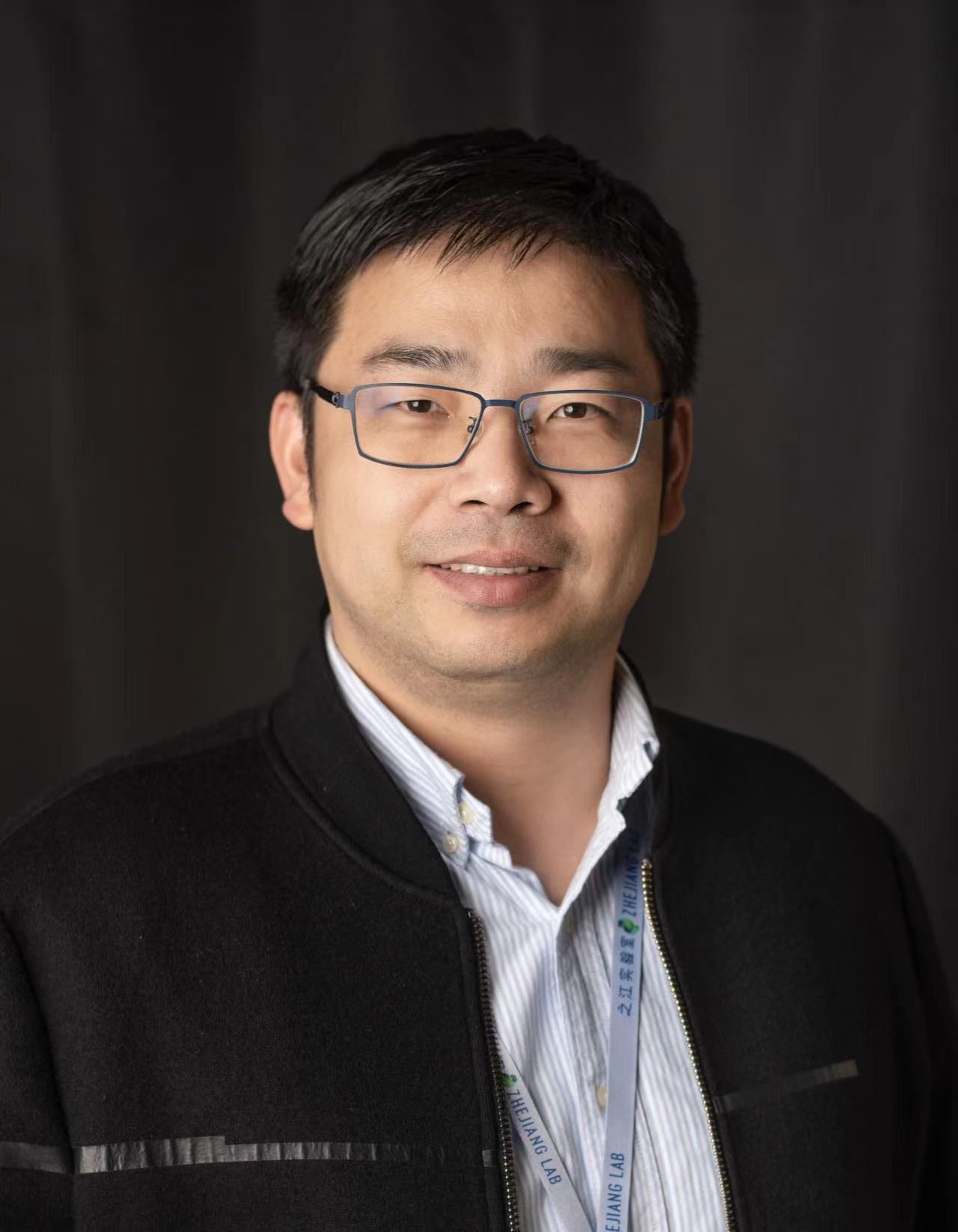}
\textbf{Chengwei Huang} (Member, IEEE) received his undergraduate degree in 2006, and Ph.D.\ for speech emotion recognition from Southeast University (China) in 2013. His main research interests include affective computing, signal processing, and digital health. He conducted research on human-computer interaction as an associate professor in Soochow University from 2013 to 2014. He is currently with the Zhejiang Laboratory as a principal investigator studying digital health, learning, and well-being.
\endbio
\vskip6pc

\bio{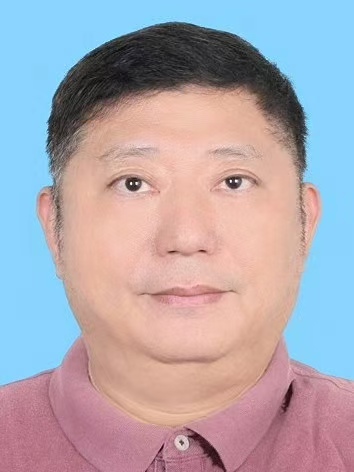}
\textbf{Cairong Zou} received the B.S., M.S., and Ph.D.\ degrees from Southeast University, China,
 in 1984, 1987, and 1991, respectively, in electrical engineering. He was a Postdoctoral Fellow with the Department of Electrical and Computer Engineering, Concordia University, Canada, in 1992. He is currently a Professor with Southeast University.
\endbio
\vskip6pc

\bio{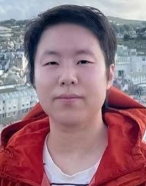}
\textbf{Wei Zhou } (IEEE Senior Member) is an Assistant Professor at Cardiff University, United Kingdom. Previously, Wei studied and worked at other institutions such as the University of Waterloo (Canada), the National Institute of Informatics (Japan), the University of Science and Technology of China, Intel, Microsoft Research, and Alibaba Group. Dr Zhou is now an Associate Editor of IEEE Transactions on Neural Networks and Learning Systems (TNNLS), ACM Transactions on Multimedia Computing, Communications, and Applications (TOMM), and Pattern Recognition. Wei’s research interests span multimedia computing, perceptual image processing, and computational vision.
\endbio

\vskip6pc

\bio{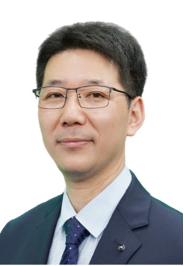}
\textbf{Weiping Ding} (M’16-SM’19) received the Ph.D. degree in Computer Science, Nanjing University of Aeronautics and Astronautics, Nanjing, China, in 2013. From 2014 to 2015, he was a Postdoctoral Researcher at the Brain Research Center, National Chiao Tung University, Hsinchu, Taiwan, China. In 2016, he was a Visiting Scholar at National University of Singapore, Singapore. From 2017 to 2018, he was a Visiting Professor at University of Technology Sydney, Australia. Now he is the Full Professor of Nantong University. His main research directions involve granular data mining and multimodal machine learning. He has published over 380 articles, including over 190 IEEE Transactions papers. His twenty authored/co-authored papers have been selected as ESI Highly Cited Papers. He serves as an Associate Editor/Area Editor/Editorial Board member of more than 10 international prestigious journals, such as IEEE Transactions on Neural Networks and Learning Systems, IEEE Transactions on Fuzzy Systems, IEEE Transactions on Intelligent Transportation Systems, Information Fusion, Neurocomputing, Applied Soft Computing, et al. He was the Leading Guest Editor of Special Issues in several prestigious journals, including IEEE Transactions on Evolutionary Computation, IEEE Transactions on Fuzzy Systems, et al. 
\endbio
\vskip6pc

\bio{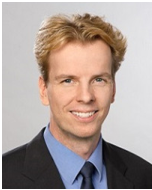}
\textbf{Bj\"orn W.\ Schuller} (Fellow, IEEE) received the diploma, the doctoral degree, and the habilitation and Adjunct Teaching Professorship in the subject area of signal processing and machine intelligence, all in electrical engineering and information technology from the Technical University of Munich (TUM), Germany, in 1999, 2006, and 2012, respectively. He is a Full Professor of artificial intelligence, and the Head of GLAM -- the Group on Language, Audio \& Music, Imperial College London, U.K., a Full Professor and Chair of Health Informatics at TUM, co-founding CEO and current CSO of audEERING. He is also with Munich's MCML, MDSI, and MIBE. Previous stations include the University of Augsburg and Passau, Germany, as Full Professor, the French CNRS, and Joanneum Research in Graz, Austria. He is  President Emeritus and Fellow of the AAAC, Fellow of the ACM, BCS, DIRDI, ELLIS, IEEE, and ISCA. He authored or coauthored five books and more than 1500 publications in peer reviewed books, journals, and conference proceedings leading to more than 65k citations (h-index = 116).
\endbio
\vskip6pc

\end{document}